\documentclass[useAMS,usenatbib]{mn2e}
\usepackage{graphicx,graphics}
\usepackage{amsmath}
\usepackage{hyperref}
\usepackage{bm}

\newcommand{\code}{\textsc{sdar}}
\newcommand{\codeurl}{https://github.com/lwang-astro/SDAR}
\newcommand{\codelink}{\href{\codeurl}{GitHub}}

\newcommand{\wv}{\bm{w}}
\newcommand{\wvz}{\bm{w}(0)}
\newcommand{\Wv}{\bm{W}}

\newcommand{\ma}{m_1}
\newcommand{\mb}{m_2}
\newcommand{\mc}{m_3}
\newcommand{\mi}{m_i}
\newcommand{\mim}{m_{3-i}}
\newcommand{\mj}{m_j}

\newcommand{\va}{\bm{v}_1}
\newcommand{\vb}{\bm{v}_2}
\newcommand{\vc}{\bm{v}_3}
\newcommand{\vi}{\bm{v}_i}

\newcommand{\vio}{\bm{v'}_i}
\newcommand{\vcm}{\bm{v}_\mathrm{cm}}

\newcommand{\ra}{\bm{r}_1}
\newcommand{\rb}{\bm{r}_2}
\newcommand{\rc}{\bm{r}_3}
\newcommand{\ri}{\bm{r}_i}
\newcommand{\rim}{\bm{r}_{3-i}}
\newcommand{\rio}{\bm{r'}_i}
\newcommand{\rvo}{\bm{r'}}
\newcommand{\rj}{\bm{r}_j}
\newcommand{\rcm}{\bm{r}_\mathrm{cm}}

\newcommand{\mab}{\ma + \mb}

\newcommand{\rccm}{\rc - \rcm}
\newcommand{\rab}{\ra - \rb}
\newcommand{\rac}{\ra - \rc}
\newcommand{\rbc}{\rb - \rc}
\newcommand{\rij}{\ri - \rj}
\newcommand{\riim}{\ri - \rim}
\newcommand{\ric}{\ri - \rc}
\newcommand{\ricmave}{\langle \ric \rangle}

\newcommand{\vccm}{\vc - \vcm}
\newcommand{\vicmave}{\langle \vi - \vcm \rangle}

\newcommand{\diff}{\mathrm{d}}

\newcommand{\Hsd}{H_\mathrm{sd}}
\newcommand{\intUj}{\mathcal{V}_{j}}
\newcommand{\intUz}{\mathcal{V}_{1}}
\newcommand{\dvi}{\delta \vi}
\newcommand{\dvio}{\delta \vio}
\newcommand{\np}{N_{\mathrm{p}}}
\newcommand{\Psd}{P_{\mathrm{sd}}}
\newcommand{\Porg}{P}
\newcommand{\Lv}{\bm{L}}

\newcommand{\Lin}{\mathcal{L}_{\mathrm{in}}}
\newcommand{\Lout}{\mathcal{L}_{\mathrm{out}}}
\newcommand{\lin}{\ell_{\mathrm{in}}}
\newcommand{\lout}{\ell_{\mathrm{out}}}

\newcommand{\Hbi}{H_{\mathrm{b},i}}
\newcommand{\Hbin}{H_{\mathrm{b,in}}}
\newcommand{\Hbout}{H_{\mathrm{b,out}}}
\newcommand{\Hpert}{H_{\mathrm{pert}}}
\newcommand{\mai}{m_{1,i}}
\newcommand{\mbi}{m_{2,i}}
\newcommand{\rki}{\bm{r}_{k,i}}
\newcommand{\rkmi}{\bm{r}_{3-k,i}}
\newcommand{\rai}{\bm{r}_{1,i}}
\newcommand{\rbi}{\bm{r}_{2,i}}
\newcommand{\vai}{\bm{v}_{1,i}}
\newcommand{\vbi}{\bm{v}_{2,i}}
\newcommand{\vki}{\bm{v}_{k,i}}

\newcommand{\mkmi}{m_{3-k,i}}
\newcommand{\vci}{\bm{v}_{\mathrm{cm},i}}
\newcommand{\mabi}{\mai + \mbi}
\newcommand{\rabi}{\rai - \rbi}
\newcommand{\nb}{N_{\mathrm{b}}}

\newcommand{\Lb}{\bm{L}_{\mathrm{b}}}
\newcommand{\Lc}{\bm{L}_{3}}
\newcommand{\Hb}{H_{\mathrm{b}}}
\newcommand{\Hi}{H_{i}}
\newcommand{\Hc}{H_{3}}
\newcommand{\ki}{\kappa_{i}}

\newcommand{\Pv}{\bm{P}}
\newcommand{\pt}{p_{\mathrm{t}}}
\newcommand{\pvi}{\bm{p}_{i}}
\newcommand{\pv}{\bm{p}}

\newcommand{\pci}{\bm{p}_{\mathrm{cm},i}}
\newcommand{\rv}{\bm{r}}
\newcommand{\vv}{\bm{v}}
\newcommand{\Rv}{\bm{R}}

\newcommand{\Tsd}{T_{\mathrm{sd}}}
\newcommand{\Usd}{U_{\mathrm{sd}}}
\newcommand{\Gsd}{\Gamma_{\mathrm{sd}}}
\newcommand{\Ti}{T_{i}}

\newcommand{\Tbi}{T_{\mathrm{b},i}}
\newcommand{\Tbcmi}{T_{\mathrm{b,cm},i}}
\newcommand{\Ubi}{U_{\mathrm{b},i}}

\newcommand{\ttl}{\textsc{ar}}
\newcommand{\hermite}{\textsc{hermite}}
\newcommand{\kepler}{\textsc{kepler}}

\newcommand{\kref}{k_{\mathrm{ref}}}

\newcommand{\lzb}{L_{\mathrm{z'},2}}

\newcommand{\mfacn}{\mathcal{M}_{n}}
\newcommand{\Pn}{P_{n}}

\newcommand{\incin}{\theta_{\mathrm{in}}}
\newcommand{\mpr}{m_{\mathrm p}}
\newcommand{\Pout}{P_{\mathrm out}}
\newcommand{\Pin}{P_{\mathrm in}}
\newcommand{\ms}{m_{\mathrm s}}
\newcommand{\tkl}{t_{\mathrm{KL}}}
\newcommand{\ein}{e_{\mathrm{in}}}
\newcommand{\ain}{a_{\mathrm{in}}}
\newcommand{\aout}{a_{\mathrm{out}}}
\newcommand{\eout}{e_{\mathrm{out}}}
\newcommand{\rin}{\bm{r}_{\mathrm{in}}}
\newcommand{\rout}{\bm{r}_{\mathrm{out}}}
\newcommand{\Eout}{E_{\mathrm{out}}}
\newcommand{\Einb}{E_{\mathrm{in2}}}
\newcommand{\aia}{a_{\mathrm{in1}}}
\newcommand{\aib}{a_{\mathrm{in2}}}
\newcommand{\eia}{e_{\mathrm{in1}}}
\newcommand{\kia}{\kappa_{\mathrm{in1}}}
\newcommand{\kib}{\kappa_{\mathrm{in2}}}
\newcommand{\tklib}{t_{\mathrm{KL,in2}}}

\newcommand{\eh}{\varepsilon(H)}
\newcommand{\ehsd}{\varepsilon(\Hsd)}
\newcommand{\egsd}{\varepsilon(\Gsd)}
\newcommand{\eg}{\varepsilon(\Gamma)}
\newcommand{\elv}{\widetilde \varepsilon(\Lv)}
\newcommand{\eli}{\widetilde \varepsilon(L_{i})}

\title[SDAR integrator]{A slow-down time-transformed symplectic integrator for solving the few-body problem}
\author[Long Wang et al.]{Long Wang$^{1,2}$\thanks{E-mail:long.wang@astron.s.u-tokyo.ac.jp}, Keigo Nitadori$^{2}$ and Junichiro Makino$^{2}$\\
  $^{1}$Department of Astronomy, School of Science, The University of Tokyo, 7-3-1 Hongo, Bunkyo-ku, Tokyo, 113-0033, Japan \\
  $^{2}$RIKEN Center for Computational Science, 7-1-26 Minatojima-minami-machi, Chuo-ku, Kobe, Hyogo 650-0047, Japan\\
}
\begin{document}

\date{Accepted --  . Received --; in original form --}

\pagerange{\pageref{firstpage}--\pageref{lastpage}} \pubyear{2002}

\maketitle

\label{firstpage}

\begin{abstract}
  An accurate and efficient method dealing with the few-body dynamics is important for simulating collisional $N$-body systems like star clusters and to follow the formation and evolution of compact binaries.
  We describe such a method which combines the time-transformed explicit symplectic integrator \citep{Preto1999,Mikkola1999} and the slow-down method \citep{Mikkola1996}.
  The former conserves the Hamiltonian and the angular momentum for a long-term evolution, while the latter significantly reduces the computational cost for a weakly perturbed binary.
  In this work, the Hamilton equations of this algorithm are analyzed in detail.
  We mathematically and numerically show that it can correctly reproduce the secular evolution like the orbit averaged method and also well conserve the angular momentum.
  For a weakly perturbed binary, the method is possible to provide a few order of magnitude faster performance than the classical algorithm.
  A publicly available code written in the c++ language, \code, is available on \codelink.
  It can be used either as a standalone tool or a library to be plugged in other $N$-body codes.
  The high precision of the floating point to $62$ digits is also supported.
\end{abstract}


\section{Introduction}

Few-body dynamics embedded in $N$-body systems like star clusters is important for many research topics:
the formation of high-velocity stars via the ejections by interacting with binaries or massive black holes \citep[e.g.][]{Poveda1967,Leonard1988,Leonard1990,Yu2003,Gualandris2005,Gvaramadze2009,Fujii2011};
the mergers of binaries that produce exotic objects like blue stragglers \citep[e.g.][]{Bailyn1995,Davies2004,Hurley2005,Heggie2008,Leigh2011,Hypki2013} and gravitational wave sources \citep[e.g.][]{PZ2000,Oleary2006,Banerjee2010,Antonini2016,Askar2017,DiCarlo2019};
and the energy generation source that controls the dynamical evolution of star clusters \citep[e.g.][]{Henon1961,Heggie1975,Hills1975,Spitzer1987,Binney1987,Breen2013,Wang2020}.

However, few-body systems are also challenging to handle in the numerical simulations.
Firstly, they can have a much shorter dynamical timescale than that of the host environment.
For example, an open cluster with a few thousand stars has a typical dynamical (crossing) time of few Myrs, but a close binary can have a period of few days.
It is very time consuming to accurately follow the lifetime of the cluster with a time resolution less than the binary period.
Secondly, the numerical error introduced by the integrator can accumulate after many binary orbits and cause an artificial drift of the orbital parameters.

To avoid these issues, the $N$-body codes for simulating star clusters, such as \textsc{nbody6} \citep{Aarseth2003}, artificially ``freeze'' the orbits of weakly perturbed binaries and hierarchical systems, i.e.,
if the perturbation is below a threshold and the hierarchical systems satisfy a stability criterion, the internal motion of the systems are not evolved until the perturbation becomes strong.
Such trick can significantly reduce the computational cost, but it ignores the long-term effect of weak perturbation.
In particular, the evolution of angular momentum, and thus the eccentricity, can be completely wrong.
Besides, whether the instability of hierarchical systems is properly treated depends on the quality of the stability criterion.

There are two approximate solutions that can properly handle the few-body systems and also avoid the time consuming calculation.
The first method is using the orbit averaged Hamiltonian.
In the case of a hierarchical triple, instead of tracing the positions and velocities of individual components by integrating the classical Newtonian equation of motion, the orbits of inner and outer binaries can be evolved by using the Hamilton equation written in the Delaunay's elements.
The short-period terms in the Hamiltonian can be eliminated (orbit averaged) by the Von Zeipel transformation \citep[e.g.][]{Naoz2013,Naoz2016}.
For example, the orbit averaged method with a quadruple-level perturbation term is provided in \cite{Naoz2013}.
Because the timescale of secular evolution is longer than the periods of binaries, such method can be very efficient to integrate the orbital evolution of a stable hierarchical triple.

Another method is called ``slow-down'' introduced by \cite{Mikkola1996}.
For a perturbed binary, by artificially slowing down the orbital motion (scaling the time) while keeping the orbital parameters unchanged, the external perturbation is effectively enlarged.
In such case, the effect of perturbation on one binary orbit can represent the average effect of several orbits.
When the perturbation is weak, this method can properly approximate the secular evolution.
The benefit is rich: it is easy to be implemented in a $N$-body code; 
can be applied for a general few-body system with an arbitrary number of binaries and singles;
is not only for a kepler binary, but also for any type of periodic orbits.
In \cite{Mikkola1996}, a constant slow-down factor is tested for a binary with a post-Newtonian type of perturbation.
However, it is not well discussed and tested how we can change the slow-down factor, even though it is crucial for the actual use of the slow-down method.
Especially, this is important for dealing with the common few-body systems in a star cluster, such as triples with high-eccentric or hyperbolic orbits of perturbers.
When the perturbation to a binary becomes smaller, we start from no slow-down to a larger slow-down factor.
But how to change it properly is not well understood.
\cite{Mikkola1996} discuss a factor-of-two change, but such a way would cause the non-conservative change in the total energy of the system.

These two methods not only can reduce the computing cost, but also avoid the large accumulate error due to much less integration steps.
But as the price of approximation, both methods lose the phase information of the orbits, thus the mean motion resonance cannot be correctly reproduced.
They also have different disadvantages.
The orbit averaged method was used to study a limited type of stable hierarchical systems, e.g. stable hierarchical triples and the Lunar system.
For a more general case like the hyperbolic encounters, systems with more than three bodies, or strongly perturbed few-body systems, to derive the equation of motion with high-order perturbation terms is unpractical.
The slow-down method is much more flexible but it cannot avoid the issue of the singularity of Newtonian force, i.e., to integrate the motion of a high-eccentric binary, a large numerical error appears at the peri-center unless the integration step is significantly reduced.

The solution for the issue of singularity is to apply the regularization algorithm.
The key idea of regularization is to remove the singularity by a transformation of the equation of motion.
The Burdet-Heggie regularization \citep{Burdet1967,Burdet1968,Heggie1973} for solving the Kepler problem uses a time-transformation, $\diff t = r \diff \tau$, where $\tau$ is a fictitious time.
By using the eccentric vector in the equation of motion together, the singularity term can be eliminated.
\cite{KS1965} introduce the KS regularization method, which transforms the equation of motion of the Kepler orbit to a form of harmonic oscillator without singularity.
\cite{Mikkola1996} show how to combine the KS regularization together with the slow-down method to have an efficient solution for integrating a few-body system.
\cite{Mikkola1999} and \cite{Preto1999} introduce the time-transformed explicit symplectic integrator (TSI) or ``Algorithmic regularization'' (AR), which can be used for any potential that only depends on the coordinates of particles.

Here we describe the TSI (AR) method in a bit more detail.
The symplectic integrator can conserve the Hamiltonian and the angular momentum of a system.
Thus it is very suitable for simulating the long-term evolution of a system.
However, it requires a constant integration step.
In the classical symplectic method, time step, $\diff t$, is also the integration step, $\diff s$.
This leads to a low efficiency in integrating an eccentric Kepler orbit.
In order to be accurately enough, $\diff t$ has to be fixed to the smallest value determined at the pericenter.
%
The solution is to apply a time transformation, $\diff t=g \diff s$, which decouples $\diff s$ and $\diff t$ with a function $g$ \citep[e.g][]{Hairer1997}.
Thus, $\diff t$ can vary to avoid the issue of low efficiency while $\diff s$ is fixed to keep the symplectic property.
This method can be described by the extended phase space Hamiltonian, where $t$ is treated as a coordinate and need to be integrated.
The disadvantage is that the time transformation usually results in an inseparable Hamiltonian and only the expensive implicit integrator can be used.
\cite{Mikkola1999} and \cite{Preto1999} find a solution by defining a specific type of $g$ (see Section~\ref{sec:AR}) that the Hamiltonian can be written in a separable style, in order to use the explicit symplectic integrator.
Especially, for an isolated binary system, if $\diff t = \diff s/ |U|$, where $|U|$ is the absolute value of the binary potential energy (similar like the Burdet-Heggie time transformation), the integrator behaviors dramatically well for the Kepler orbit, i.e., the numerical trajectory follows the exact one with a phase error of time.

In this work, we develop an efficient method for simulating few-body systems by combining the slow-down and the TSI (AR) schemes.
In Section~\ref{sec:Hsd}, we show the slow-down Hamiltonian and the equation of motion for several types of few-body systems, such as perturbed binaries, triples and general hierarchical systems.
We demonstrate that the slow-down method can correctly reproduce the secular evolution and conserve the angular momentum.
In Section~\ref{sec:AR}, the TSI (AR) method is described in detail.
The combination of two, the slow-down time-transformed symplectic method, is discussed in Section~\ref{sec:gsd}.
The implementation and numerical tests are shown in Section~\ref{sec:implementation} and \ref{sec:test}.
Finally, we discuss our results and draw conclusions in Section~\ref{sec:conclusion}.

To be convenient, hereafter $\rv$, $\pv$, $\vv$ and $m$ represent coordinates, conjugate momenta, velocities and masses of particles separately.
We define the phase-space vector as $\wv\equiv (\rv, \pv)$.
The suffix of a variable using an index (e.g. $i$, $j$, $1$, $2$ ...) represents a specific particle while no suffix represents all particles in a $N$-body system.
For exmample, $\wv \equiv $($\wv_1$, $\wv_2$,..., $\wv_N$).

\section{Slow-down Hamiltonian}
\label{sec:Hsd}

The slow-down method can be described by a modified Hamiltonian \citep{Mikkola1996}.
For a perturbed binary, the slow-down Hamiltonian is 
\begin{equation}
  \Hsd =  \frac{1}{\kappa} \Hb + (H - \Hb), 
  \label{eq:sd}
\end{equation}
where $H$ is the original Hamiltonian of the system, $\Hb$ is the Hamiltonian of the binary and $\kappa$ is the slow-down factor.
The equation of motion is
\begin{equation}
  \begin{aligned}
    \frac{\diff \wv}{\diff t} = \{ \wv, \Hsd \},
  \end{aligned}
  \label{eq:dH}
\end{equation}
where $\{\}$ is the Possion bracket.
When $\kappa = 1$, $\Hsd \equiv H$.
The important point is that although the Hamiltonian is modified, $\wv$ keeps the original definition.
Thus, the slow-down affects the time evolution of orbital phase, but the positions and velocities of particles are not scaled.

\subsection{Perturbed binary system}
\label{sec:bin}

We first consider a simple example of one binary with an external potential $U(\rv)$ and a fixed $\kappa$,
\begin{equation}
  \Hsd =  \frac{1}{\kappa} \left [ \frac{1}{2} \ma \va^2 + \frac{1}{2} \mb \vb^2 - \frac{G \ma \mb}{|\rab|} \right ] + U(\rv).
\end{equation}
where the suffixes ``1'' and ``2'' denote the two components of the binary, and $\pvi$ is replaced by $\mi \vi$.

Applying Eq.~\ref{eq:dH}, the evolution of $\rv$ and $\vv$ can be described by
\begin{equation}
  \begin{aligned}
    \frac{\diff \vi}{\diff t} = & -\frac{1}{\kappa} \frac{G \mim (\riim)}{|\riim|^3} - \frac{\partial U(\rv)}{\partial \ri} & (i=1~\mathrm{or}~2)\\
    \frac{\diff \ri}{\diff t} = & \frac{1}{\kappa} \vi. \\
  \end{aligned}
  \label{eq:H2}
\end{equation}
Thus, in the reference frame of the perturber, the motion of binary is slowed down by a factor of $\kappa$, while in the view of the binary, the perturbation is enhanced by $\kappa$ times.
We can understand this better by considering the cumulative effect of perturbation to $\vi$ on one slow-down orbit.

The effective period of the slow-down binary has $\Psd = \kappa \Porg$, where $\Porg$ is the original period.
By integrating $\diff \vi /\diff t$ of Eq.~\ref{eq:H2}, the change of $\vi$ after one $\Psd$ is
\begin{equation}
  \dvi = \int_0^{\Psd} {\frac{\diff \vi }{\diff t}}.
  \label{eq:dviH2}
\end{equation}
In the case of no perturbation, $\vi(t=\Psd) = \vi(t=0)$.
Thus
\begin{equation}
  \int_{0}^{\Psd} {\frac{\diff \vi}{\diff t}} = 0.
\end{equation}
When the effect of $U(\rv)$ is weak,
\begin{equation}
  \dvi \approx \int_{0}^{\Psd} {- \frac{\partial U(\rv)}{\partial \ri}}.
\end{equation}
Then $\dvi$ is only determined by the integrated effect of perturbation.
Here $\ri$ only passes one cycle, but the integration time is $\kappa \Porg$.
Thus, the effect the perturbation is amplified by $\kappa$ times on one orbit of binary.

On the other hand, when $\kappa$ is an integer, we can measure $\dvi$ in the original case (no slow-down; denoted as $\dvio$) after the same physical time by a summation of $\kappa$ pieces of integral, where each piece represents one orbit:
\begin{equation}
  \begin{aligned}
  \dvio & = \sum_{j=1}^{\kappa} \int_{(j-1) \Porg}^{j \Porg} {- \frac{\partial U(\rvo)}{\partial \rio}}\\
        & = \sum_{j=1}^{\kappa} \intUj, \\
  \end{aligned}
\end{equation}
If the perturbation does not change significantly,
\begin{equation}
  \begin{aligned}
    \intUj \approx \intUz & &(j = 1, \kappa).
  \end{aligned}
  \label{eq:UieqUz}
\end{equation}
This means the effect of perturbation can be represented by the integration of one orbit ($\intUz$) multiplied by $\kappa$ times.
Thus, it is equivalent to the treatment of the slow-down method and $\dvi \approx \dvio$.
In other words, the slow-down method approximates the perturbation in an orbit averaged way.
However, when the external potential changes significantly within the time interval of $\kappa P$, Eq.~\ref{eq:UieqUz} becomes invalid.
Thus, there is a threshold of $\kappa$.
We discuss the criterion in Section~\ref{sec:varykappa}.

\subsection{Triple system}

For a hierarchical triple, 
\begin{equation}
  \begin{aligned}
    \Hsd = & \frac{1}{\kappa}\left [ \frac{1}{2} \ma (\va-\vcm)^2 + \frac{1}{2} \mb (\vb-\vcm)^2 - \frac{G \ma \mb}{|\rab|} \right ] \\
     + & \frac{1}{2} (\mab) \vcm^2 + \frac{1}{2} \mc \vc^2 - \frac{G \ma \mc}{|\rac|} - \frac{G \mb \mc}{|\rbc|}, 
  \end{aligned}
  \label{eq:H3}
\end{equation}
where the suffixes ``1'' and ``2'' denote the two components of the inner binary and ``3'' represents the third body, and $\vcm$ is the center-of-the-mass velocity of the inner binary.
Because the slow-down affects only the internal motion of the binary in its rest frame, $\vcm$ is subtracted from the slow-down term.

Applying Eq.~\ref{eq:dH} (with a fixed $\kappa$ to Eq.~\ref{eq:H3}), the evolution of $\rv$ and $\vv$ can be described by
\begin{itemize}
\item {\it binary components} ($i = 1~\mathrm{or}~2$):
  \begin{equation}
    \begin{aligned}
      \frac{\diff \vi}{\diff t} = &-\frac{1}{\kappa} \frac{G \mim (\riim)}{|\riim|^3} - \frac{G \mc (\ric)}{|\ric|^3} \\
      \frac{\diff \ri}{\diff t} = &~\frac{1}{\kappa} (\vi -\vcm) + \vcm; \\
    \end{aligned}
    \label{eq:eom3b}
  \end{equation}
\item {\it third body} ($i=3$):
  \begin{equation}
    \begin{aligned}
      \frac{\diff \vi}{\diff t} = & -\sum_{j=1}^{2}\frac{G \mj (\rij)}{|\rij|^3} \\
      \frac{\diff \ri}{\diff t} = & ~\vi. \\
    \end{aligned}
    \label{eq:eom3s}
  \end{equation}
\end{itemize}
The form of the third body is identical to the original one.

\begin{figure}
  \centering
  \includegraphics[width=0.85\columnwidth]{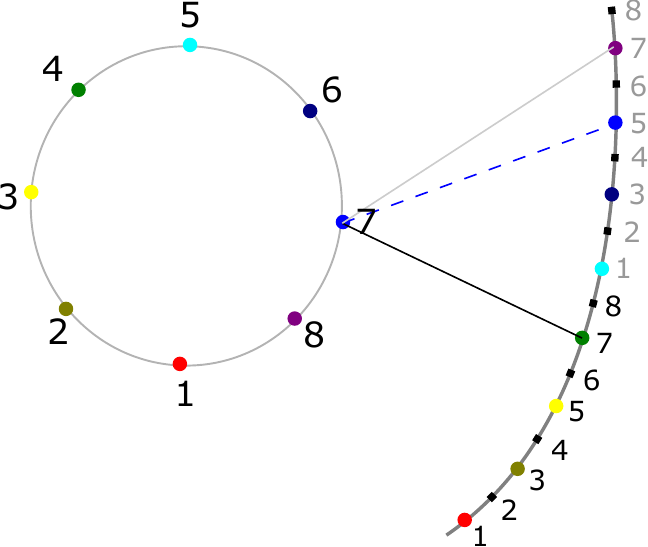}
  \caption{The illustration to show in a hierarchical triple how the force from the perturber to the binary component is calculated in the slow-down and the original methods.
    The left circle represents the orbit of one binary component and the right curve indicates the track of the perturber.
    The points alone the two tracks indicate the corresponding positions (same color or number) for calculating the perturbation force.
    The number represents the original method where binary pass two cycles while the color represents the slow-dowm method with $\kappa=2$ (one cycle).
    For example, the two solid lines and one dashed line link the corresponding positions (dark blue point with number of $7$) in the original and the slow-down cases separately.
  }
  \label{fig:sdsample}
\end{figure}

We demonstrate how the perturbation force is calculated in the slow-down and the original integration methods in Fig.~\ref{fig:sdsample}.
Along the binary orbit (one of the component), $8$ sample points with an equal angular separation are chosen for the demonstration.
When the binary finish two cycles in the original case, each sample point has twice force calculations from the perturber at the corresponding positions shown with the same number along its track.
In the slow-down case with $\kappa=2$, the binary finish one orbit when the perturber passes the same track.
Thus, the corresponding position of the perturber (with same colors) for each sample point is different from that of the original case.
If the sample points indicate the step sizes in an integrator, the slow-down method reduces the step sizes by $\kappa$ times.

The two types of lines that connect the dark blue point $7$ represent one example of the perturbation calculation in the two methods separately.
If the solid and dashed lines have very close length and direction, the slow-down method give a good approximation of the perturbation force.
This is the case when the perturber is far away ($|\rc| \gg |\rab|$), i.e., the perturbation is weak.

\subsubsection{Orbit averaged Hamiltonian}

For a hierarchical triple, we can also describe the slow-down method in terms of the orbit averaged method.
By using the Legendre expansion for the perturbation term \citep[e.g.][]{Harrington1968,Naoz2013,Naoz2016}, $\Hsd$ can be written as the combination of three components:
\begin{equation}
  \begin{aligned}
    \Hsd = & \frac{1}{\kappa} \Hbin + \Hbout + \Hpert \\
    \Hbin = & - \frac{G \ma \mb }{2 \ain}\\
    \Hbout = &  - \frac{G (\mab) \mc}{2 \aout} \\
    \Hpert = &- \frac{G}{|\rout|} \sum_{n=2}^{\infty} \mfacn \left (\frac{|\rin|}{|\rout|} \right )^{n} \Pn (\cos{\Psi}),\\
  \end{aligned}
  \label{eq:Hsdorb}
\end{equation}
where $\Pn$ is the Legendre polynomials, the suffixes ``in'' and ``out'' indicate the inner and outer binaries, $a$ represents the semi-major axis, $\rin = \rb - \ra$, $\rout = \rc - \rcm$, $\rcm$ is the center-of-the-mass position of the inner binary and $\Psi$ is the angle between $\rin$ and $\rout$.
The mass coefficient sequence,
\begin{equation}
  \mfacn = \ma \mb \mc \frac{\ma^{n-1} - (-\mb)^{n-1}}{(\mab)^{n}}.
\end{equation}

Using the Delaunay's elements, the two binary terms of Hamiltonian have the form
\begin{equation}
  \begin{aligned}
    \Hbin  = & - \frac{G^2 \ma^3 \mb^3 }{2 (\mab) \Lin^2} \\
    \Hbout = & - \frac{G^2 (\mab)^3 \mc^3}{2 (\mab + \mc) \Lout^2} \\
  \end{aligned}
\end{equation}
where the conjugate momenta, 
\begin{equation}
  \begin{aligned}
    \Lin = & \frac{\ma \mb }{\mab} \sqrt{G (\mab) \ain}\\
    \Lout= & \frac{(\mab) \mc}{\mab + \mc} \sqrt{G (\mab +\mc) \aout}.\\
  \end{aligned}
  \label{eq:L}
\end{equation}
Applying the equation of motion, the time derivations of their corresponding coordinates, $\lin$ and $\lout$, are
\begin{equation}
  \begin{aligned}
    \frac{\diff \lin}{\diff t} = & \frac{\partial \Hsd}{\partial \Lin} 
                               =  \frac{1}{\kappa} \frac{G^2 \ma^3 \mb^3 }{(\mab) \Lin^3} + \frac{\partial \Hpert}{\partial \Lin}\\
    \frac{\diff \lout}{\diff t} = & \frac{\partial \Hsd}{\partial \Lout} 
                               =  \frac{G^2 (\mab)^3 \mc^3 }{(\mab+\mc) \Lout^3} + \frac{\partial \Hpert}{\partial \Lout}.\\
  \end{aligned}
  \label{eq:lorb}
\end{equation}

The effect of slow-down is reflected on the first term of $\diff \lin/\diff t$, which represents the mean motion of the inner binary.
Without the perturbation term, Eq.~\ref{eq:lorb} indicates that the mean motion of the inner binary is $\kappa$ times slower of that in the original case, as discussed in Section~\ref{sec:bin}.
The orbit averaged method uses a canonical transformation (Von Zeipel transformation) to remove the short-period terms ($\lin$ and $\lout$) in $\Hpert$\citep[e.g.][]{Naoz2013}:
\begin{equation}
  \overline{\Hpert}   = \int_{0}^{2\pi}{\int_{0}^{2\pi}{\Hpert \diff \lin \diff \lout}}.
  \label{eq:Hpertintl}
\end{equation}
Since $\lin$ and $\lout$ do not appear in $\Hbin$ and $\Hbout$ and $\Hpert$ is independent of (the fixed) $\kappa$, the transformation results in the same form of $\overline{\Hpert}$ in the slow-down and original cases.
Thus, the slow-down method provides the same secular evolution as the orbit averaged way.

\subsection{$N$-body system with multiple slow-down binaries}

For a general few-body system containing $\nb$ binaries, each binary can have an individual slow-down factor, $\ki$.
The slow-down Hamiltonian has the general form:
\begin{equation}
  \begin{aligned}
    \Hsd = & \sum_{i}^{\nb} \frac{1}{\ki} \Hbi + \left ( H - \sum_{i} \Hbi \right )\\
    \Hbi = & \frac{1}{2} \mai (\vai-\vci)^2 + \frac{1}{2} \mbi (\vbi-\vci)^2 \\
           & - \frac{G \mai \mbi}{|\rabi|} \\
    \vci = & \frac{\mai \vai + \mbi \vbi}{\mabi}, \\
  \end{aligned}
  \label{eq:Hsdg}
\end{equation}
where the suffixes ``$1,i$'' and ``$2,i$'' represent the two components in the $i^{\mathrm{th}}$ binary.
To be convenient, we use the first $2\nb$ indices to indicate binary components and others to indicate singles, i.e., ``1,i'' and ``2,i'' are equivalent to $2i-1$ and $2i$ separately.

Applying Eq.~\ref{eq:dH} with all $\ki$ being fixed, the equation of motion is
\begin{itemize}
\item {\it binary components} ($i = 1,2,...,\nb; k = 1~\mathrm{or}~2$):
  \begin{equation}
    \begin{aligned}
      \frac{\diff \vki}{\diff t} = & - \frac{1}{\ki} \frac{G \mkmi (\rki-\rkmi)}{|\rki-\rkmi|^3} \\
                                   & - \sum_{j=1;j\ne 2i+1-k}^{N}\frac{G \mj (\rki-\rj)}{|\rki-\rj|^3} \\
      \frac{\diff \rki}{\diff t} = &~\frac{1}{\ki} (\vki -\vci) + \vci; \\
    \end{aligned}
    \label{eq:eomNb}
  \end{equation}
\item {\it single body} ($i=2\nb+1,2\nb+2,...,N$):
  \begin{equation}
    \begin{aligned}
      \frac{\diff \vi}{\diff t} = & -\sum_{j=1;j\ne i}^{N}\frac{G \mj (\rij)}{|\rij|^3} \\
      \frac{\diff \ri}{\diff t} = & ~\vi. \\
    \end{aligned}
    \label{eq:eomNs}
  \end{equation}
\end{itemize}
This general form suggests that $\ki$ only explicitly affects the internal motion of the binary $i$.
The force calculations of singles and other binaries do not explicitly depend on $\ki$.

\subsection{Varying $\kappa$ in the integration}
\label{sec:varykappa}

In previous sections, we assume that $\kappa$ is constant.
In many few-body systems, the perturbation to a binary can vary significantly, especially when a nearby perturber has a highly eccentric or hyperbolic orbit.
Thus, varying $\kappa$ during the integration is necessary to balance the performance and accuracy.
\cite{Mikkola1996} suggests to calculate $\kappa$ based on the ratio between the tidal force from perturbers and the internal force of the binary assuming the separation is $2 \ain$.
For a system with one binary and $\np$ perturbers, we define a modified version by including the eccentricity ($e$),
\begin{equation}
    \kappa(\wv) = \kref \frac{\ma \mb}{(\mab) \left [\ain(1+\ein) \right ]^{3}} \sum_{i}^{\np} \frac{|\ri-\rcm|^3}{\mi} ,
  \label{eq:kappa}
\end{equation}
where $\kref$ is a constant coefficient.

Since this $\kappa$ depends on $\wv$, one might think that the additional terms of ($\Hbin \partial \kappa(\wv)/\partial \rv$, $\Hbin \partial \kappa(\wv)/\partial \vv$) should appear in the equation of motion.
However, these terms actually should not be included to ensure a correct secular evolution.
%
In the case of a triple system, by applying Eq.~\ref{eq:kappa} in the Hamiltonian form of Eq.~\ref{eq:Hsdorb}, 
\begin{equation}
  \frac{1}{\kappa} \Hbin = - \frac{\kref G (\mab) \mc}{2 |\rout|} \left [\frac{\ain(1+\ein)}{|\rout|} \right ]^2.
\end{equation}
$\Hbin$ becomes to depend on $\lout$ (from $\ain/|\rout|$) in the same order as that of the perturbation term in $\Hpert$ with $n=2$ .
The additional terms including $\Hbin$ then affects the orbital average of $\lout$ (Eq.~\ref{eq:Hpertintl}), which  may not result in the same secular evolution as the orbit averaged method.

On the other hand, when Eq.~\ref{eq:UieqUz} is satisfied, $\kappa(\wv)$ is expected to evolve slowly in one step.
As the first order approximation, $\kappa$ can be treated as a temporary constant and is only updated at the end of one integration step. Essentially, $\kappa$ is only a step function of time.
Then the unwarranted effect can be avoided.
We can describe this Jumping-$\kappa$ method for a system with $\nb$ binaries as:
\begin{enumerate}
\item advance $\diff t$ with fixed $\ki$ measured at $t$.
\item update $\ki$ at the new time $t' = t + \diff t$;
\item apply an instant correction of $\Hsd$ by
  \begin{equation}
    \delta \Hsd(t') = \sum_{i}^{\nb} {\left ( \frac{1}{\ki(t')} - \frac{1}{\ki(t)} \right) \Hbi(t')}.
    \label{eq:dHsd}
  \end{equation}
\end{enumerate}
With the step (iii), even $\Hsd$ is not conserved, the cumulative numerical error of $\Hsd$ can still be properly tracked as
\begin{equation}
  \epsilon(\Hsd,k \diff t) = \Hsd(k \diff t) - \sum_{i=1}^{k} \delta \Hsd(i \diff t) - \Hsd(0),
  \label{eq:ehsd}
\end{equation}
where $k$ is the total step count.

In this method, although the unwarranted effect on the secular evolution is not included in one integration step, the instant change of $\Hsd$ may still introduce an additional effect.
Thus, it is necessary to limit the change rate of $\kappa$ for safety.
We estimate how $\kappa(\wv)$ changes depending on the orbital parameters.
With Eq.~\ref{eq:kappa}, the differential of $1/\kappa$,
\begin{equation}
  \begin{aligned}
    \diff \left( \frac{1}{\kappa} \right) = &\kref \sum_{i}^{\np} \frac{3 (\mab) \mi}{\ma \mb} \frac{a^3}{|\ri-\rcm|^3} \\
    &\left( \frac{\diff a}{a} - \frac{\diff |\ri-\rcm|}{|\ri-\rcm|}\right). \\
  \end{aligned}
  \label{eq:dkappa}
\end{equation}
Typically, in the weak perturbation condition, $a$ does not change significantly but $|\ri-\rcm|$ may vary a lot if the perturber has an eccentric or hyperbolic orbit.
In the case of one perturber, Eq.~\ref{eq:dkappa} can be simplified as
\begin{equation}
  \diff \left( \frac{1}{\kappa} \right) \approx - 3 \kref \left( \frac{1}{\kappa} \right) \frac{\diff |\rccm|}{|\rccm|}.
  \label{eq:dkappasim}
\end{equation}
Thus, the time derivation of $1/\kappa$ depends on $|\vccm|/|\rccm|$.
The evolution timescale is the order of the crossing time of the perturber.

The slow change of $\kappa$ requires that the strength of perturbation does not vary significantly within at least one $\Psd$.
Based on Eq.~\ref{eq:dkappasim}, we can introduce a timescale criterion to limit the maximum value of $\kappa$ for safety:
\begin{equation}
  \kappa_{\mathrm{max}} =  \frac{c |\ricmave|}{ \Psd|\vicmave|},
  \label{eq:kmaxt}
\end{equation}
where $c$ is a constant coefficient; $\ricmave$ and $\vicmave$ represent the mass weighted average of velocities and positions of all perturbers referring to the center-of-the-mass of the binary separately.
If a binary is the perturber, the mass weights help to remove the velocity fluctuations caused by the internal motion of the two components of the perturber.
For the Jumping-$\kappa$ method, Eq.~\ref{eq:kappa} and \ref{eq:kmaxt} together provide the criterion to estimate $\kappa$.
In Section~\ref{sec:test}, we show that the numerical tests by using the Jumping-$\kappa$ method indeed provide a correct secular evolution.

\subsection{Conserved quantities}
\subsubsection{Energy}
\label{sec:Esd}

It is important to know, with the irregular form of $\Hsd$, what are conserved quantities during the evolution that can be used to check the quality of the integration.
When $\kappa$ is a constant, $\Hsd$ does not explicitly depend on time.
Thus the ``slow-down energy'', $\Hsd$, is a conserved quantity, which means that the physical energy, $H$, changes during the evolution.
The variation of $H$ can be calculated by
\begin{equation}
  \delta H = H- \Hsd =\sum_{i}^{\nb} {\left (1-\frac{1}{\ki} \right ) \Hbi}.
  \label{eq:diffHsd}
\end{equation}
Thus, $H$ oscillates in a timescale of the secular evolution of the binaries.

When $\kappa$ is not fixed, whether $\Hsd$ is conserved depends on how $\kappa$ is evaluated.
If $\kappa$ follows Eq.~\ref{eq:kappa} exactly and the equation of motion contains the additional terms of ($\Hbin \partial \kappa(\wv)/\partial \rv$, $\Hbin \partial \kappa(\wv)/\partial \vv$), $\kappa$ would not explicitly depend on time, thus $\Hsd$ is conserved.
However, this also indicates that a significant variation of $\kappa$ results in a large change of binary energy.
Such behaviour is introduced by the unwarranted effect discussed in Section~\ref{sec:varykappa}. 
Instead, the Jumping-$\kappa$ method that exclude the additional term can avoid such problem, although $\Hsd$ explicitly depends on time and is not conserved.
Since the physical energy is anyway not conserved, it is more important to ensure the correct behaviour of secular evolution rather than to conserve $\Hsd$.
On the other hand, we can still follow the integration error by using Eq.~\ref{eq:ehsd} in the absence of the energy conservation.

\subsubsection{Angular momentum}
\label{sec:L}

The angular momentum defined by
\begin{equation}
  \Lv = \sum_{i} \mi \ri \times \vi
\end{equation}
is a conserved quantity in the original case since 
\begin{equation}
  \{\Lv, H \} = \bm{0},
\end{equation}
where $\bm{0}$ is a zero vector.
Interestingly, it can be proved that $\Lv$ is also conserved with $\Hsd$ (Eq.~\ref{eq:Hsdg}; see Appendix), i.e.
\begin{equation}
  \{\Lv, \Hsd \} = \bm{0}.
  \label{eq:LHsd}
\end{equation}
This conservation is not only valid for an invariant $\kappa$.
With the help of a symbolic calculator \footnote{We use \textsc{python.sympy} to prove Eq.~\ref{eq:LHsd}.}, it can be shown that for $\kappa$ defined in Eq.~\ref{eq:kappa}, Eq.~\ref{eq:LHsd} is also satisfied.
On the other hand, if $\kappa$ explicitly depends on time, $\Lv$ is still conserved because the differential in Eq.~\ref{eq:LHsd} is independent of time.
Thus, $\Lv$ is generally conserved in the slow-down method and can be used to validate the quality of the integration.

\section{Time transformed symplectic (AR) method}
\label{sec:AR}

The TSI (AR) method is accurate for solving the long-term evolution of few-body systems.
It can be described by the extended phase-space Hamiltonian \citep[e.g.][]{Hairer1997,Preto1999}:
\begin{equation}
  \Gamma(\Wv) = g(\Wv) [H(\wv,t) - H(\wvz,0)],
  \label{eq:gamma}
\end{equation}
where $H(\wv,t)$ is the standard Hamiltonian and $g(\Wv)$ is time-transformation function.
The extended phase-space vector, $\Wv=(\Rv,\Pv)$, is defined by including an additional pair of the coordinate, $t$, and the conjugate momentum, $\pt = - H(\wvz,0)$ (negative initial energy).
$\wvz$ is the initial value of $\wv$.


With the new differential variable $s$, 
\begin{equation}
  g(\Wv) = \frac{\diff t}{\diff s}.
\end{equation}
A special type of $g(\Wv)$ introduced by \cite{Preto1999},
\begin{equation}
  g(\Wv) = \frac{f(T(\Pv)) - f(-U(\Rv))}{T(\Pv) + U(\Rv)} ,
  \label{eq:g}
\end{equation}
leads to a separable $\Gamma$:
\begin{equation}
  \Gamma(\Wv) = f(T(\Pv)) - f(-U(\Rv)),
  \label{eq:gammas}
\end{equation}
where the kinetic energy, $T(\Pv) \equiv T(\pv) + \pt$, and the potential energy, $U(\Rv) \equiv U(\rv,t)$.

Putting $\Gamma(\Wv)$ in the equation of motion,
\begin{equation}
  \frac{\diff \Wv}{\diff s} = \{ \Wv, \Gamma(\Wv) \},
\end{equation}
the derivatives of $\Wv$ with respect to $s$ are
\begin{equation}
  \begin{split}
    \frac{\diff \ri }{\diff s} &= f'(T(\pv)+\pt) \frac{\partial T(\pv)}{\partial \pvi} \\
    \frac{\diff  t  }{\diff s} &= f'(T(\pv)+\pt)\\
    \frac{\diff \pvi}{\diff s} &= f'(-U(\rv,t)) \frac{\partial U(\rv,t)}{\partial \ri} \\
    \frac{\diff \pt }{\diff s} &= f'(-U(\rv,t)) \frac{\partial U(\rv,t)}{\partial t}.\\
  \end{split}
  \label{eq:eomg}
\end{equation}

\begin{figure}
  \centering
  \includegraphics[width=1.0\columnwidth]{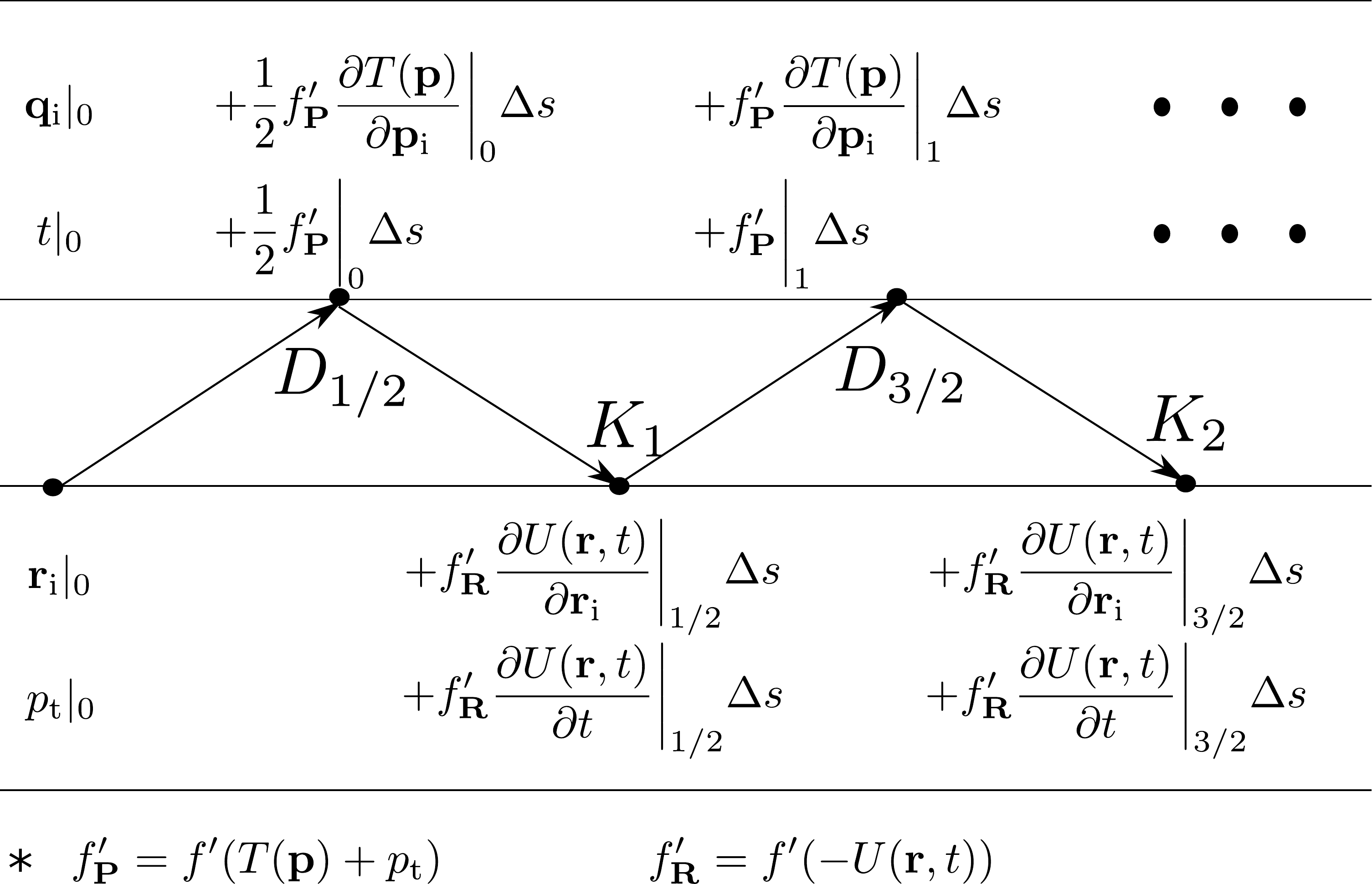}
  \caption{The schedule of the LogH method with the equation of motion shown in Eq.~\ref{eq:eomg} and ~\ref{eq:logf}.
  }
  \label{fig:extH}
\end{figure}

\cite{Preto1999} and \cite{Mikkola1999} showed that by choosing
\begin{equation}
  f(x) = \log{x},
  \label{eq:logf}
\end{equation}
the Leap-Frog method with drift-kick-drift (DKD) mode can ensure that the numerical trajectory of a Kepler orbit follows the exact one, $\wv(t)$, with only a phase error of time.
Hereafter we use ``LogH'' to represent this algorithm.
If the time error is not important, the LogH method is very efficient to integrate a weakly perturbed Kepler orbit.
Fig.~\ref{fig:extH} shows the schedule of one DKD loop.

\begin{figure}
  \centering
  \includegraphics[width=1.0\columnwidth]{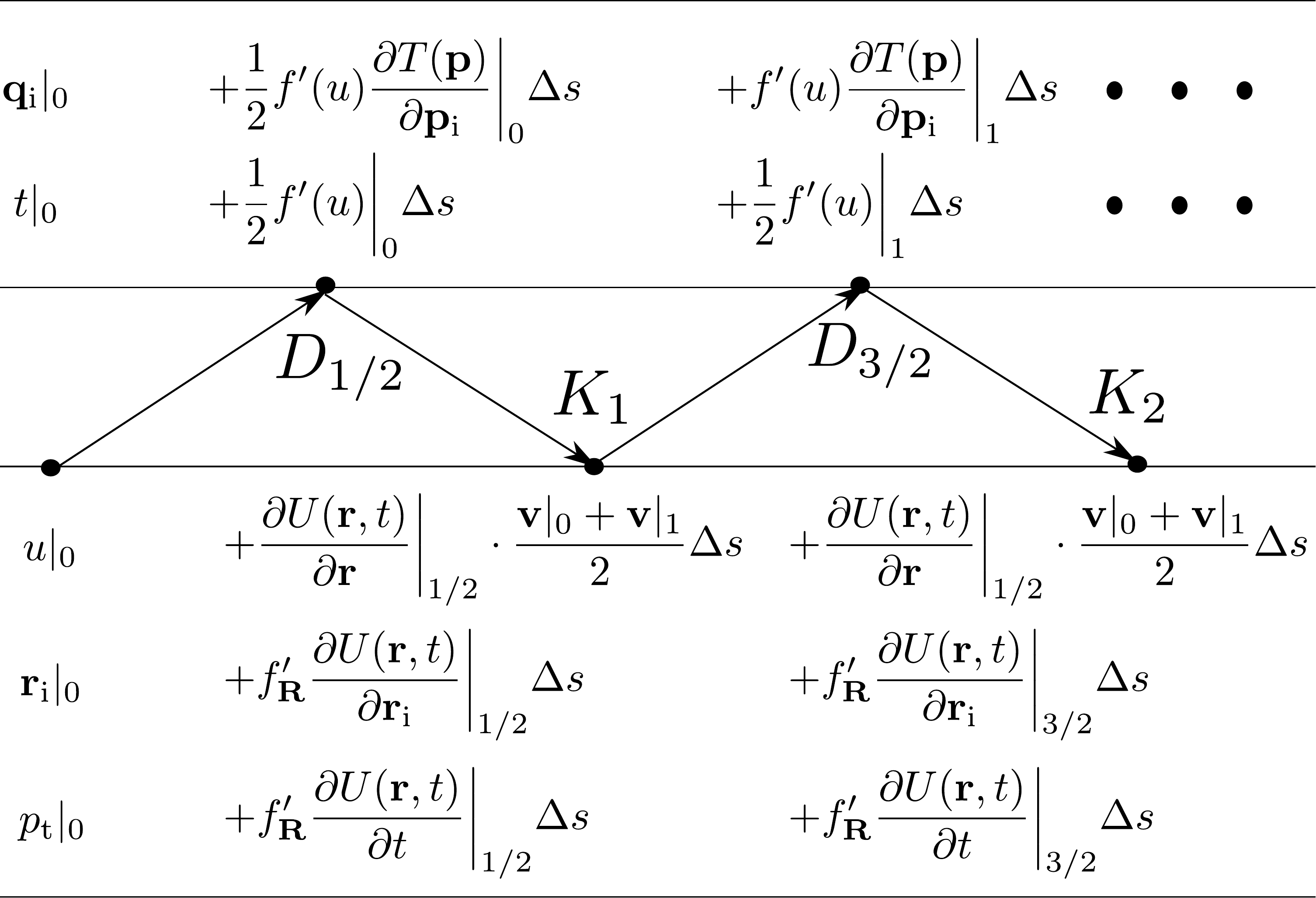}
  \caption{The schedule of the Mikkola (2012) scheme.}
  \label{fig:extW}
\end{figure}

When $H$ does not explicitly depends on $t$, $H(\wv,t) = H(\wvz,0)$, and $T(\Pv)+U(\Rv) = 0$.
Thus $f'(T(\Pv))=f'(-U(\Rv))$.
\cite{Mikkola2002} found that instead of calculating $T(\Pv)$, one can also define a variable,
\begin{equation}
  u = \int \frac{\partial U(\Rv)}{\partial \Rv} \cdot \frac{\diff \Rv}{\diff t}
\end{equation}
Then $f'(u) = f'(T(\Pv))$.
The integration schedule of this method for one DKD step is shown in Fig.~\ref{fig:extW}.

\subsection{Geometric feature}

Via the Taylor expansion, Eq.~\ref{eq:g} can be approximated as \citep{Preto1999}
\begin{equation}
  g(\Wv) \approx f'(-U(\Rv))
  \label{eq:gapr}
\end{equation}
By applying Eq.~\ref{eq:logf} for a Kepler orbit system, the relation between $t$ and $s$ is
\begin{equation}
  \diff t \approx \frac{\diff s}{-U(\Rv)} = \frac{|\ra-\rb|\diff s}{G \ma \mb}.
  \label{eq:rds}
\end{equation}
This is equivalent to the case in Burdet-Heggie regularization method \citep{Burdet1967,Burdet1968,Heggie1973}.
It can be shown that this time transformation can remove the singularity by including the energy and the eccentric vector in the equation of motion.
This suggests that the LogH method can well handle the high-eccentric orbit.

On the other hand, we can show that $s$ has a geometric meaning.
For a Kepler orbit.
\begin{equation}
  |\rab| = a (1 - e \cos E)
  \label{eq:raeE}
\end{equation}
where $E$ is eccentric anomaly.
If $E=0$ indicates time zero, the relation between $E$ and $t$ is
\begin{equation}
  t = \frac{\Porg}{2\pi} (E - e \sin E).
  \label{eq:tE}
\end{equation}
The differential of this equation results in
\begin{equation}
  \diff t = \frac{\Porg}{2\pi} (1 - e \cos E) \diff E.
\end{equation}
By using Eq.~\ref{eq:raeE}, the time derivative of $E$ is
\begin{equation}
  \frac{\diff E}{\diff t} = \frac{a}{|\rab|} \frac{2\pi}{\Porg}.
\end{equation}
This suggests that $s$ represents a scaled $E$:
\begin{equation}
  \diff s = \ma \mb \sqrt{\frac{G a}{\mab}} \diff E = \mathcal{L} \diff E,
  \label{eq:dsLdE}
\end{equation}
Thus, the fixed $\diff s$ in the symplectic integrator indicates a constant step of $\diff E$.
Since $t$ is unknown before the integration finish, it is difficult to determine $\diff s$ to stop the integration at a given $t$.
However, the geometric meaning of $s$ provides a way to stop at a certain $E$. 
This is very useful for determining the number of steps per orbit in order to control the numerical accuracy of the integration.

\subsection{Conservation of energy}

Although the LogH method ensures that the numerical trajectory follows the exact Kepler orbit, the numerical $H(\wv)$ is not perfectly conserved.
Especially, when the orbit reaches the pericenter, the error of $H(\wv)$ is large due to the truncation of floating points.
Actually, based on the definition, the exact conserved ``energy'' is $\Gamma(\Wv)$.
With Eq.~\ref{eq:gamma}, \ref{eq:logf} and \ref{eq:gapr},
\begin{equation}
  \Gamma(\Wv) \approx \frac{|\rab|}{G \ma \mb} \left [ H(\wv,t) - H(\wvz,0) \right ].
\end{equation}
Since $|\rab|$ reaches the minimum at the pericenter, the large error of $H(\wv)$ is cancelled out in $\Gamma(\Wv)$.
In Section~\ref{sec:test}, we show the numerical test of such behaviour.
This feature indicates that we should avoid the calculation the orbital parameters or stop the integration at the pericenter in order to avoid a large numerical error of the orbital elements.

\section{Slow-Down time-transformed symplectic (SDAR) method}
\label{sec:gsd}

The slow-down method helps to solve the large timescale issue of a hierarchical system, while the TSI (AR) method can conserve $\Gamma$ and $\Lv$.
The combination of two (SDAR method) is expected to be an efficient algorithm to deal with the long-term evolution of few-body systems.
In the TSI method, $\Gamma(\Wv)$ can take any separable Hamiltonian, thus it is straight forward to implement $\Hsd(\wv,t)$ into $\Gamma(\Wv)$.
With Eq.~\ref{eq:Hsdg} and \ref{eq:gammas}, we can obtain the time transformed slow-down Hamiltonian for a $N$-body system with $\nb$ binaries:
\begin{equation}
  \begin{aligned}
    \Gsd(\Wv)  = & f(\Tsd(\Pv)) - f(-\Usd(\Rv)) \\
    \Tsd(\Pv) = & \sum_{i=1}^{\nb} {\frac{1}{\ki} \Tbi(\vai,\vbi)} + \sum_{i=1}^{\nb} \Tbcmi(\vci) \\
    & + \sum_{i=2\nb+1}^{N} \Ti(\vi)  + \pt \\
  \Usd(\Pv) = & \sum_{i=1}^{\nb} {\frac{1}{\ki} \Ubi(\rai,\rbi)} + \left [U(\rv) - \sum_{i=i}^{\nb} \Ubi(\rai,\rbi) \right ], \\
  \end{aligned}
\end{equation}
where
\begin{equation}
  \begin{aligned}
  \Tbi(\vai,\vbi) = &\frac{1}{2}\mai \left (\vai-\vci \right )^2  + \frac{1}{2}\mbi \left (\vbi-\vci \right)^2 \\
  \Tbcmi(\pci) = & \frac{1}{2} (\mabi) \vci^2\\
  \Ubi(\rai,\rbi) = &- \frac{G \mai \mbi}{|\rabi|} \\
  U(\rv) = & - \sum_{i=1}^{N} \sum_{j=i+1}^{N} \frac{G \mi \mj}{|\rij|}. \\
   \end{aligned}
   \label{eq:Gsd}
\end{equation}
The equation of motion of $\Gsd(\Wv)$ is the same as Eq.~\ref{eq:eomg} except that $T(\Pv)$ and $U(\Pv)$ are replaced by $\Tsd(\Pv)$ and $\Usd(\Pv)$ separately.

Notice that here the conserved ``energy'' is $\Gsd$ (with fixed $\kappa$).
Thus when the Jumping-$\kappa$ method (see Section~\ref{sec:varykappa}) is applied, the correction term in the third step should be modified as
\begin{equation}
  \begin{aligned}
    \delta \Gsd(t') = & f \left [\Tsd[\Pv(t'),\kappa(t')] \right] -  f \left [\Usd[\Rv(t'),\kappa(t')] \right ]  - \\
    & \left [ f \left [\Tsd[\Pv(t'),\kappa(t)] \right] -  f \left [\Usd [\Rv(t'),\kappa(t)] \right ] \right ]\\
  \end{aligned}
  \label{eq:dGsd}
\end{equation}
The corresponding cumulative numerical error, $\egsd$, can be obtained by replacing $\Hsd$ to $\Gsd$ in Eq.~\ref{eq:ehsd}.

\section{Implementation}
\label{sec:implementation}

Based on the SDAR method, we develop a software library, \code, written in the C++ language with the object oriented programming \footnote{URL: \codeurl}.
This library contains three modules: \ttl, \hermite~and \kepler.

\ttl~is the implementation of the SDAR method shown in Eq.~\ref{eq:Gsd} with the Jumping-$\kappa$ algorithm.
The high-order explicit symplectic integrator introduced by \cite{Yoshida1990} is used.
Both the LogH and the \cite{Mikkola2002} methods are implemented.

\hermite~is a hybrid integrator which combines a $4^{th}$ order block-time-step Hermite method and the \ttl~module.
The Hermite integrator is for the global $N$-body system and $\ttl$ deals with the compact subgroups (few-body systems).
The subgroup can contain inner binaries with individual $\ki$, while the system as a whole can also apply the slow-down method if it has a periodical evolution.
Thus, \hermite~allows a nested two-level slow-down treatment.
This can accelerate the integration of a weakly perturbed stable hierarchical system.

\kepler~is a tool to construct a Hierarchical (Kepler) binary tree for a group of particles.
It is used to identify the inner binaries and calculate $\kappa$.

The quadruple-double precision library, \textsc{qd} \citep{Hida2001}, is included in the code.
Thus the user can switch on the high-precision support (up to $62$-digit precision).
Notice that when \textsc{qd} is used, the performance is reduced and also it is not thread-safe.

\section{Numerical test}
\label{sec:test}

\begin{figure}
  \centering
  \includegraphics[width=0.9\columnwidth]{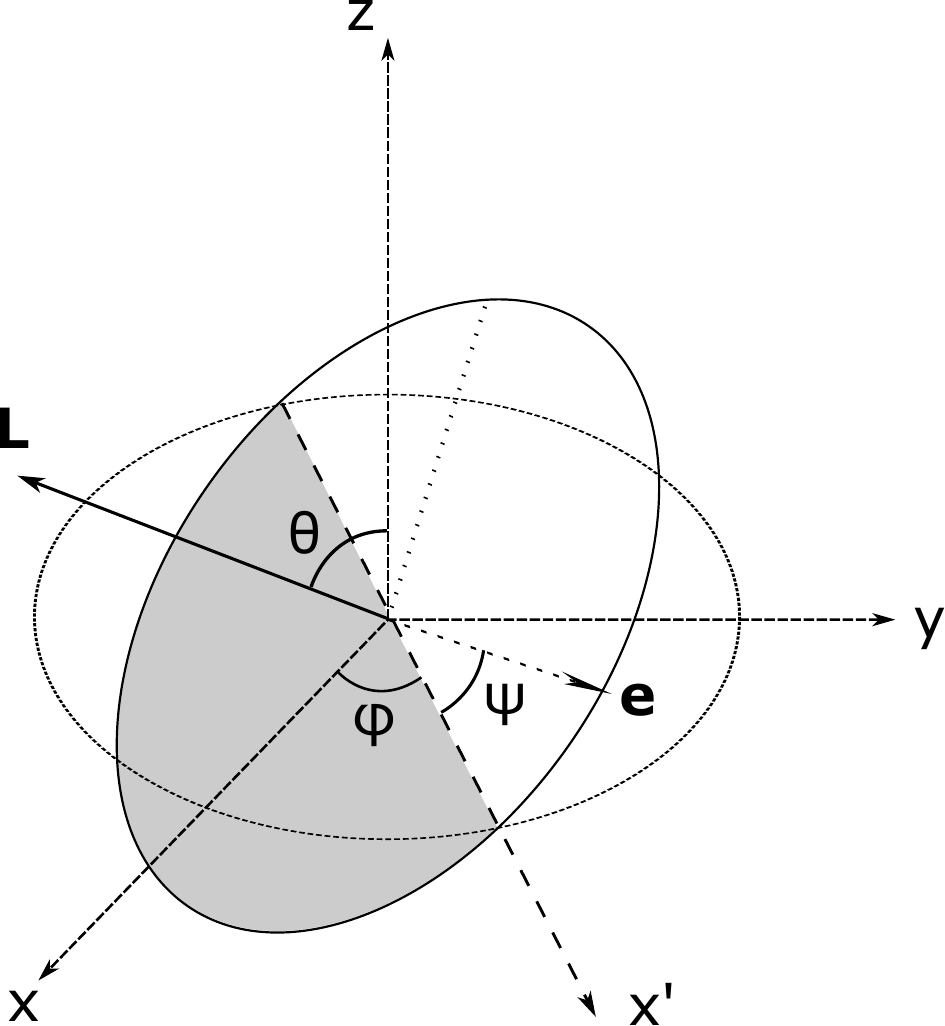}
  \caption{The Delaunay's elements (three Euler angles) of a binary orbit in the Cartesian coordinate system ($x$-$y$-$z$).
    $\theta$: inclination, $\phi$: longitude of the ascending node, and $\psi$: argument of periapsis.}
  \label{fig:euler}
\end{figure}

In this section, we check whether the slow-down method can correctly reproduce the secular evolution of few-body systems.
The comparison between the slow-down and original algorithms is done by using the \ttl~module with the $6^{th}$-order symplectic method \citep[Solution A in Table 1 from][]{Yoshida1990}.
We use the Delaunay's elements to describe the geometric information of the binary orbits.
By selecting the Cartesian coordinate system ($x$-$y$-$z$), the three angles are (see Fig.~\ref{fig:euler}):
\begin{itemize}
\item $\theta$: inclination; the angle from $z$-axis to $\Lv$.
\item $\phi$: longitude of the ascending node (often denoted as $\Omega$); the angle from $x$-axis to the intersection of the orbital plane and the $x$-$y$ plane ($x'$-axis).
\item $\psi$: argument of periapsis (often denoted as $\omega$); the angle from eccentric vector to the $x'$-axis.
\end{itemize}

\subsection{Hierarchical triple (B-S)}

The Kozai-Lidov effect is one of the most important feature appearing in a family of hierarchical triple systems \citep{Kozai1962,Lidov1962}.
For example, when the outer binary has a circular orbit and the secondary of the inner binary has a negligible mass compared to the other two bodies, the $z'$ component of the orbit averaged angular momentum of the inner binary is a conserved quantity:
\begin{equation}
  \lzb \propto \sqrt{(1-\ein^2)} \cos (\incin) = \mathrm{const},
\end{equation}
Where the $z$ axis is defined in the invariable plane.
When $\incin$ is in the range of $\cos^{-1}(\pm \sqrt{3/5})$, significant oscillations of $\ein$ and $\incin$ can happen.
In the astrophysical environment, such oscillation can result in a high eccentricity that may trigger the merger of the inner binary.
Thus, it is important to validate that the slow-down method can correctly reproduce the Kozai-Lidov effect.

We perform a simulation of a hierarchical triple (B-S).
The formula from \cite{Antognini2015} is used to estimate the Kozai-Lidov oscillation timescale:
\begin{equation}
  \tkl \approx \frac{8}{15\pi} \left ( 1 + \frac{\ma}{\mc} \right ) \frac{\Pout^{2}}{\Pin} (1 - \eout^{2})^{3/2}.
  \label{eq:tkl}
\end{equation}
For the purpose of test, parameters which allow a large $\kappa$ and a short $\tkl$ is preferred.
Thus, smaller $\ma/\mc$ and $\Pout/\Pin$ or higher eccentricity of outer binary are better.
However, the former also reduce the maximum $\kappa$ (Eq.~\ref{eq:kappa}), thus the latter is good for both requirements.

One suitable initial condition is shown in Table~\ref{tab:b-s}, where $\tkl \approx 45$.
For convenience, we use the scale-free unit (gravitational constant is one).
This is also applied for all numerical tests discussed below.

\begin{table*}
  \caption{The initial condition of the hierarchical triple (B-S) system for test. $\mp$ and $\ms$ are the masses of the primary and the secondary of inner and outer binaries. The values are shown in the scale-free unit with the gravitational constant, $G=1$.}
  \begin{tabular}{ccccc ccccc}
    \hline
          & $\mpr$ & $\ms$ & $a$ & $e$ & $\theta$ & $\phi$ & $\psi$ & $E$ & $\Porg$ \\
    \hline
    in & 0.900 & 0.100 & 0.00100 & 0.900 & 1.50 & 0.00 & 0.00 & 3.14 & $1.97\times10^{-4}$ \\
    out & 2.00 & 1.00 & 1.00 & 0.990 & 0.10 & 0.00 & 0.00 & 3.14 & 3.63 \\
    \hline
  \end{tabular}
  \label{tab:b-s}
\end{table*}

Both the inner and outer binaries are initially near the apo-centre positions ($E\approx 3.14$).
By choosing $\kref=1.0\times 10^{-6}$ as suggested by \cite{Mikkola1996} in Eq.~\ref{eq:kappa}, $\kappa_{\mathrm{max}} \approx 52$ when the outer body is at the apo-centers.
As the outer eccentricity, $\eout$, is large, $\kappa$ decreases to below one based on Eq.~\ref{eq:kappa} at the pericenter position.
This should be avoided, thus we let $\kappa\ge1.0$.
To obtain a high numerical accuracy, we use $256$ steps per orbit of the inner binary.
This results in $\Delta s\approx 6.98\times 10^{-5}$.
To reach about $4\tkl$, the total number of integration step without slow-down is about $2.36\times 10^{8}$.
In the slow-down case, it is about $3.8\times10^{7}$ steps which is $6$ times smaller.

\begin{figure}
  \centering
  \includegraphics[width=1.0\columnwidth]{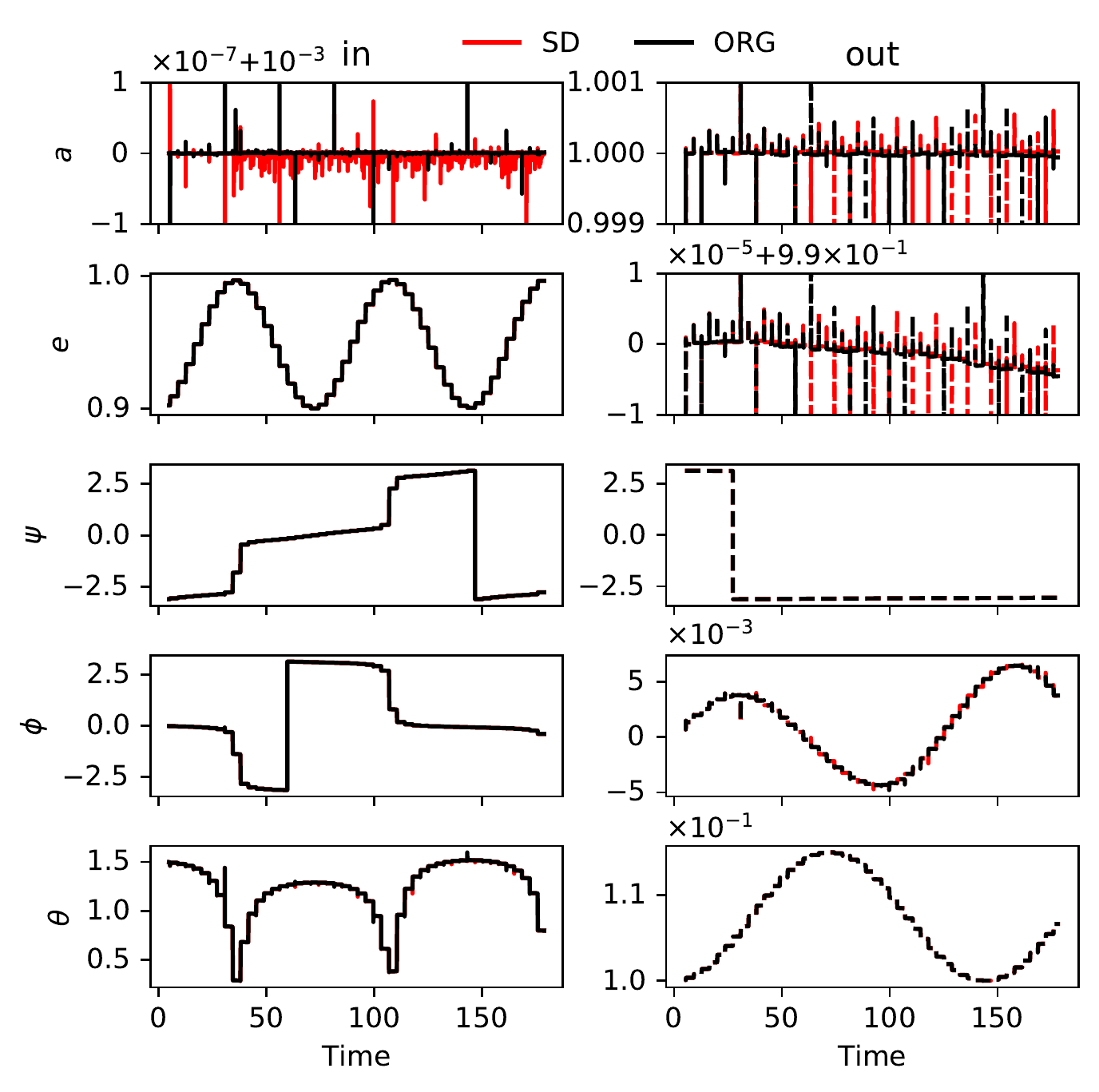}
  \caption{The evolution of orbital parameters of the inner and outer binaries for the B-S system.
    The red color represent the result using the slow-down method (SD) and the black color represents the case of no slow-down (ORG).
    For each panel, we apply the scientific notation in the plotting style of $y$-axis: the actual values of $y$-axis are calculated by $y_{\mathrm{tick}}\times scale + y_{\mathrm{offset}}$, where $y_{\mathrm{tick}}$ is the value shown along the the y-axis, $scale$ is the first value shown above the y-axis ($scale=1$ in default) and $y_{\mathrm{offset}}$ is the second value following the symbol ``+'' ($y_{\mathrm{offset}}=0$ in default). 
  }
  \label{fig:bs}
\end{figure}

Fig.~\ref{fig:bs} shows the simulation result of the B-S system by using the LogH method with and without slow-down (hereafter named as ``SD'' and ``ORG'' methods) \footnote{The scientific notation in the plotting style of $y$-axis (described in Fig.~\ref{fig:bs}) is defined by the \textsc{python} module \textsc{matplotlib.ticker} (see https://matplotlib.org/ for reference). In all figures, we follow this style in order to save space.}.
The Kozai-Lidov oscillation appears in the evolution of the orbital parameters.
The comparison between red (SD) and black (ORG) curves clearly indicates that the SD method can well reproduce the Kozai-Lidov effect with the correct timescale.
All orbital parameters, including $a$, $e$ and three Euler angles of both inner and outer binaries are consistent with each others.
The oscillation timescale is not exacted as predicted by Eq.~\ref{eq:tkl}.
Since the secondary of the inner binary has a small but none zero mass, the B-S system does not exactly satisfy the test-particle condition that is assumed in Eq.~\ref{eq:tkl}.

\begin{figure}
  \centering
  \includegraphics[width=0.8\columnwidth]{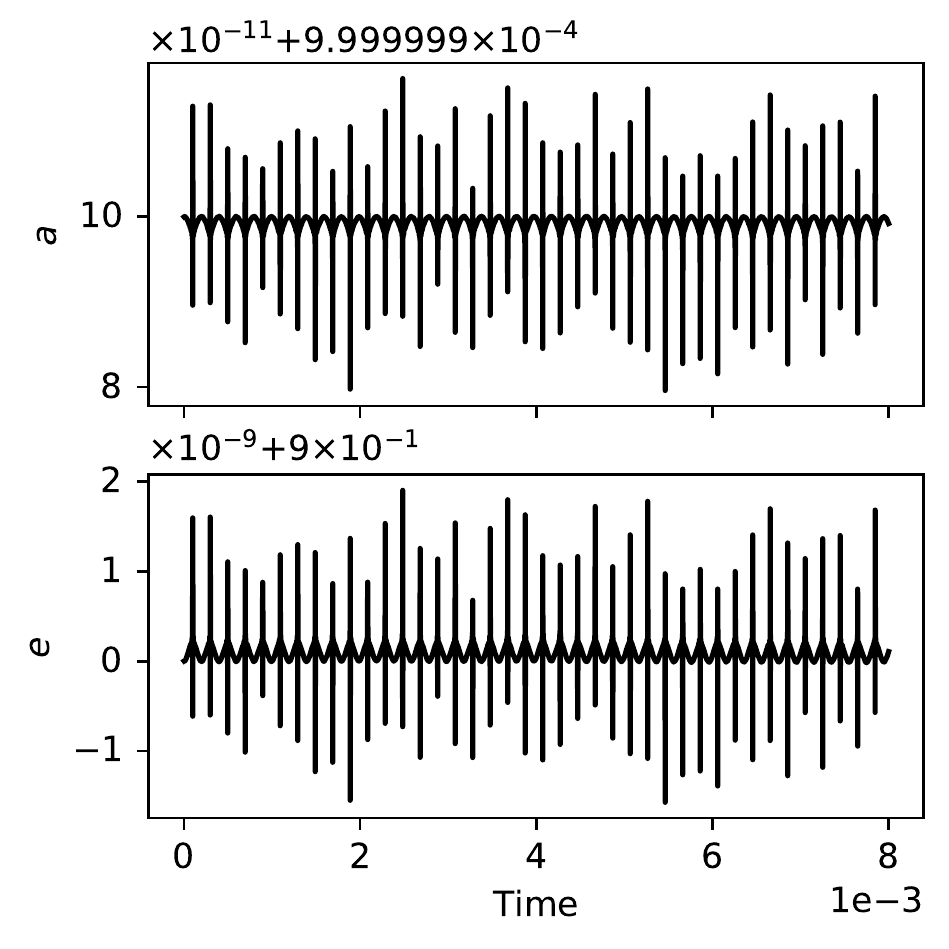}
  \caption{The evolution of $a$ and $e$ for the inner binary of the B-S system with a high-time-resolution by using the ORG method.}
  \label{fig:bshi}
\end{figure}

There are a few sharp peaks in the evolution of $a$ and $e$.
This is due to the low time resolution of the plotting (only few thousands data points along the $x$-axis).
Fig.~\ref{fig:bshi} shows the  high-time-resolution evolution of $\ain$ and $\ein$ with $t$ in $0\sim8\times 10^{-3}$ by using the ORG method.
When the inner binary passes the pericenter, sharp peaks appear.
The amplitude of peaks depends on the orbital phase of both inner and outer binaries.
Since the time interval of one peak is very short, most of the peaks are not sampled in the low-time-resolution plot and they appear by chance.
These peaks also exist in other orbital parameters but cannot be seen in the plots due to a large amplitude of the secular variation.
As the peaks do not affect the secular evolution, they can be safely ignored.

\begin{figure}
  \centering
  \includegraphics[width=0.8\columnwidth]{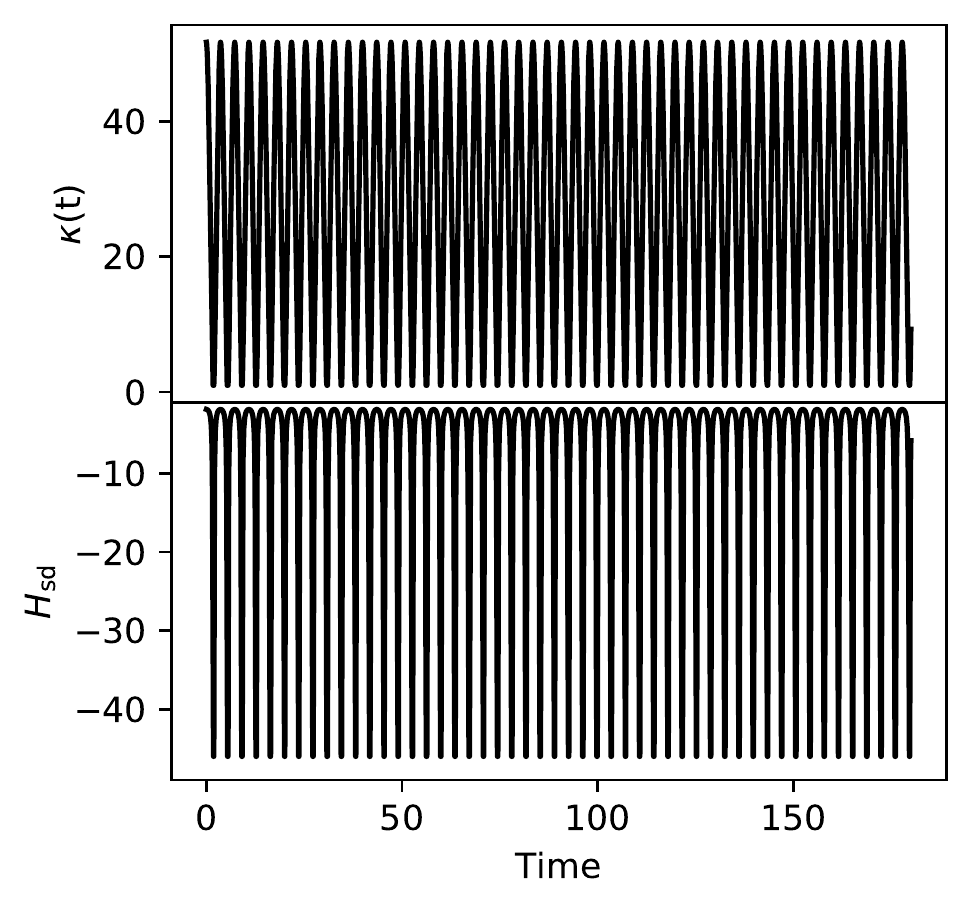}
  \caption{The evolution of $\kappa$ and $\Hsd$ for the B-S system by using the SD method.}
  \label{fig:bshsd}
\end{figure}

$\kappa$ oscillates when the third body cycles between the apo-center and the pericenter, as shown in the upper panel of Fig.~\ref{fig:bshsd}.
The evolution of $\Hsd$ follows the pattern of $\kappa$ as described by Eq.~\ref{eq:diffHsd}.
Although the scatter of $\Hsd$ is large, the behaviour of the secular evolution is correct.
This indicates that whether $\Hsd$ is conserved does not matter.

\begin{figure}
  \centering
  \includegraphics[width=0.8\columnwidth]{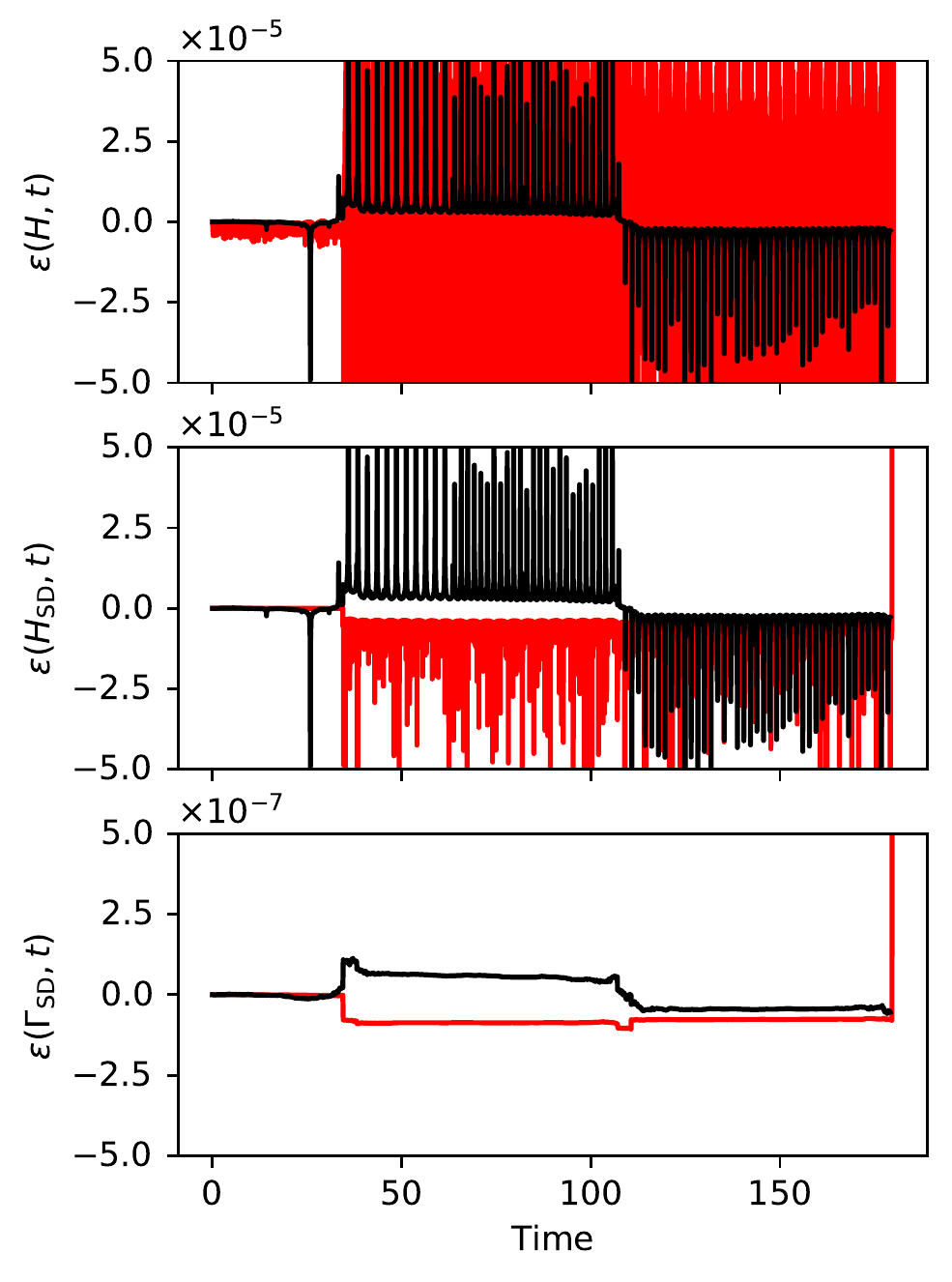}
  \caption{The evolution of $\eh$, $\ehsd$ and $\egsd$ for the B-S system.
    In the ORG method, $\eh$ and $\eg$ are equivalent to $\ehsd$ and $\egsd$ with $\kappa=1$ separately.
    To be convenient, we use $\ehsd$ and $\egsd$ as the $y$-axis labels to represent both the SD and ORG cases (a similar style is used in other plots).
    The definition of colors are the same as in Fig.~\ref{fig:bs}.
    Notice in the plot of $\egsd$, the scale of $y$-axis is two order of magnitude smaller compared to the upper plots.}
  \label{fig:bsde}
\end{figure}

However, we still can trace the numerical error by using $\egsd$.
In Fig.~\ref{fig:bsde} the evolution of the cumulative numerical errors of different definitions of Hamiltonian, $\eh$, $\ehsd$ and $\egsd$, are compared.
$\eh$ is the error of original Hamiltonian, defined as $H(t) - H(0)$.
After $\ein$ pass the maximum value for the first time, large oscillations appears on both $\eh$ and $\ehsd$.
$\eh$ of the SD method has a much larger scatter compared to the ORG method, while in the case of $\ehsd$, both have a similar amplitude.
This is consistent with the expectation as discussed in Section~\ref{sec:Esd}.
On the other hand, the scatter of $\eg$ or $\egsd$ is two order of magnitude smaller than $\ehsd$ without oscillations, which indicates that the conserved quantities are $\Gamma$ (ORG) and $\Gsd$ (SD) within one integration step.
Thus, $\egsd$ represents the numerical errors of the integrator.

\begin{figure}
  \centering
  \includegraphics[width=0.8\columnwidth]{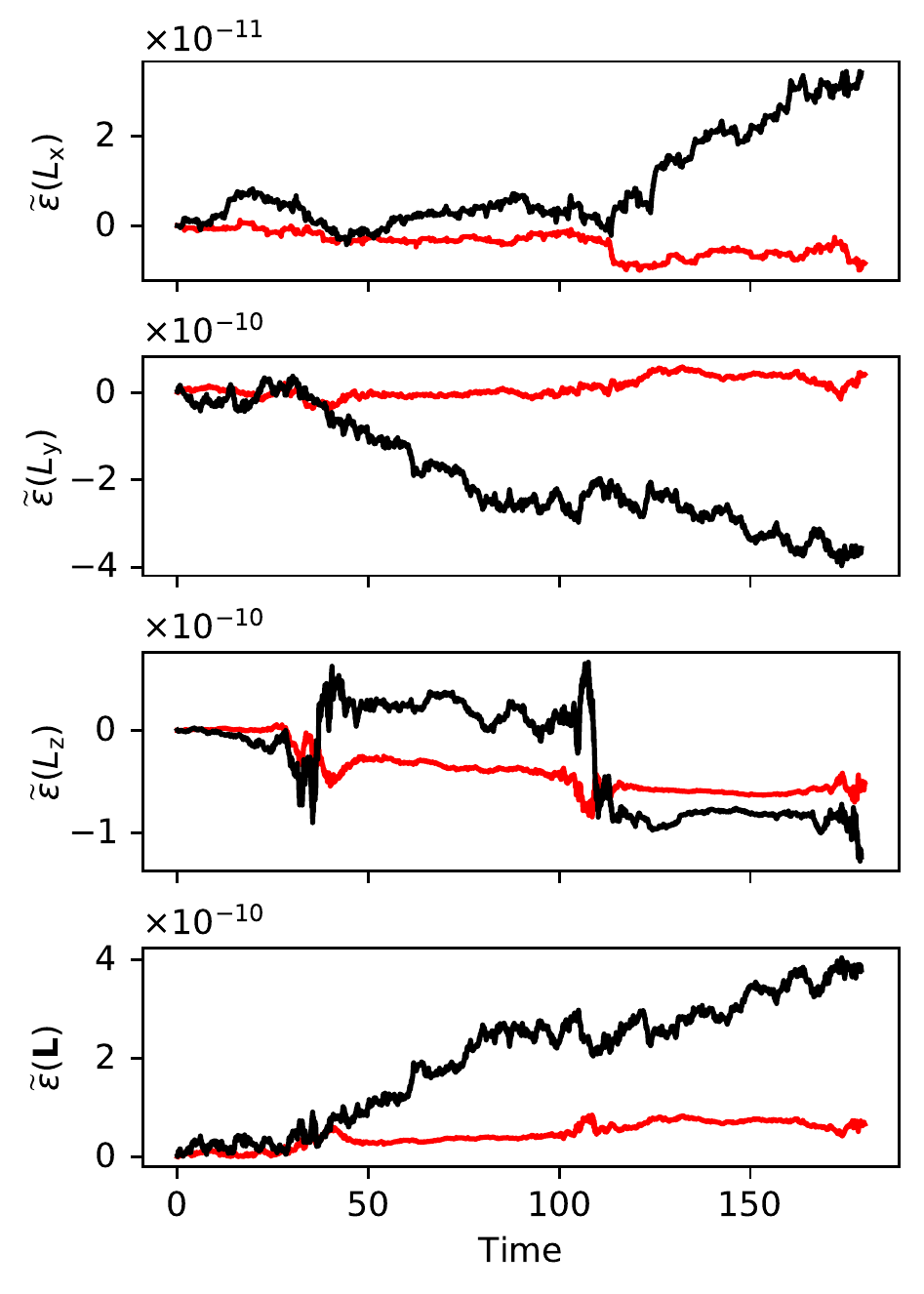}
  \caption{The cumulative error of three components of the total angular momentum of the B-S system normalized to $\Lv(0)$ (initial value). The definition of colors are the same as in Fig.~\ref{fig:bs}.}
  \label{fig:bsL}
\end{figure}

The angular momentum conservation is also checked in Fig.~\ref{fig:bsL}.
The normalized cumulative error of three components,
\begin{equation}
  \begin{aligned}
    \eli = \frac{L_{i}(t) - L_{i}(0)}{|\Lv(0)|} & & (i = x,y,z),
    \end{aligned}
\end{equation}
and the total one,
\begin{equation}
  \elv = \frac{\sqrt{\left [\Lv(t) - \Lv(0)\right ]^2}}{|\Lv(0)|},
\end{equation}
are shown.
Both the SD and the ORG methods provide a similar level of relative numerical errors ($10^{-10}$) in the three components of $\Lv$ and also in $|\Lv|$.
Besides, the error is independent of the variation of $\kappa$.
This is consistent as we proved in Section~\ref{sec:L}.

\subsection{Hierarchical quadruple (B-B)}

The mean motion resonance cannot be properly reproduced by the slow-down method.
When there are multiple binaries, the orbital resonance between inner binaries can occur if their periods are in resonance ratios.
We investigate this effect by simulating a quadruple system (B-B).
The initial condition (Table~\ref{tab:b-b}) is generated by splitting the third body of the B-S system to a binary, which has the same period as another.
Here after the suffixes ``in1'' and ``in2'' indicate the two inner binaries.
$\tkl$ of the two binaries are about $45$ and $88$.
We evolve the system to $t=160$ with the same $ds$ used in the integration of the B-S system.
The ORG method takes about $8.7\times10^{8}$ steps while the SD method takes about $9.9\times 10^{7}$ steps ($9$ times faster).

\begin{table*}
  \caption{The initial condition of the hierarchical quadruple (B-B) system.}
  \begin{tabular}{ccccc ccccc}
    \hline
          & $\mpr$ & $\ms$ & $a$ & $e$ & $\theta$ & $\phi$ & $\psi$ & $E$ & $\Porg$ \\
    \hline
    in1 & 0.900 & 0.100 & 0.00100 & 0.900 & 1.50 & 0.00 & 0.00 & 3.14 & $1.97\times10^{-4}$ \\
    in2 & 1.800 & 0.200 & 0.00126 & 0.900 & 0.10 & 0.00 & 0.00 & 3.14 & $1.97\times10^{-4}$ \\
    out & 2.00 & 1.00 & 1.00 & 0.990 & 0.100 & 0.00 & 0.00 & 3.14 & 3.63 \\
    \hline
  \end{tabular}
  \label{tab:b-b}
\end{table*}

\begin{figure*}
  \centering
  \includegraphics[width=0.8\textwidth]{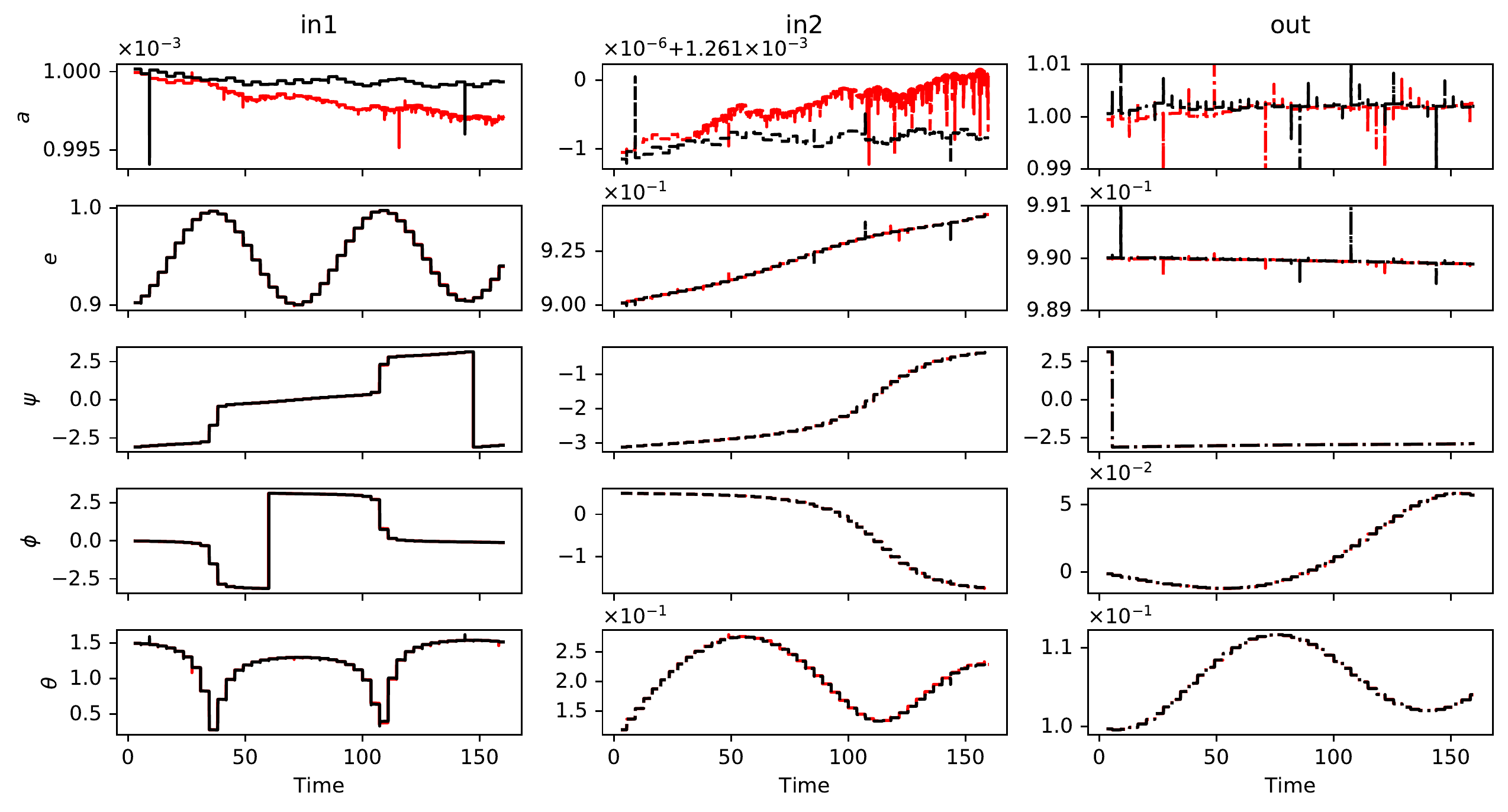}
  \caption{The evolution of orbital parameters of the two inner and outer binaries for the B-B system.
    The plotting style is similar to Fig.~\ref{fig:bs}.}
  \label{fig:bb}
\end{figure*}

The orbital evolution is shown in Fig.~\ref{fig:bb}.
$\tklib$ of the simulated result is a much larger than the prediction of Eq.~\ref{eq:tkl}.
The behaviour of the first inner binary is similar to the case of the B-S system.
Compared to the case of B-S system, the cumulative divergence of $a$ for the SD and ORG methods becomes obvious.
After about two $\tkl$, the relative difference is the order of $10^{-3}$ for $\aia$ and $\aib$.
However, this error can be neglected since the Kozai-Lidov effect dominates the secular evolution.
Thus, the SD method can still provide a reasonable result.

\begin{figure}
  \centering
  \includegraphics[width=0.8\columnwidth]{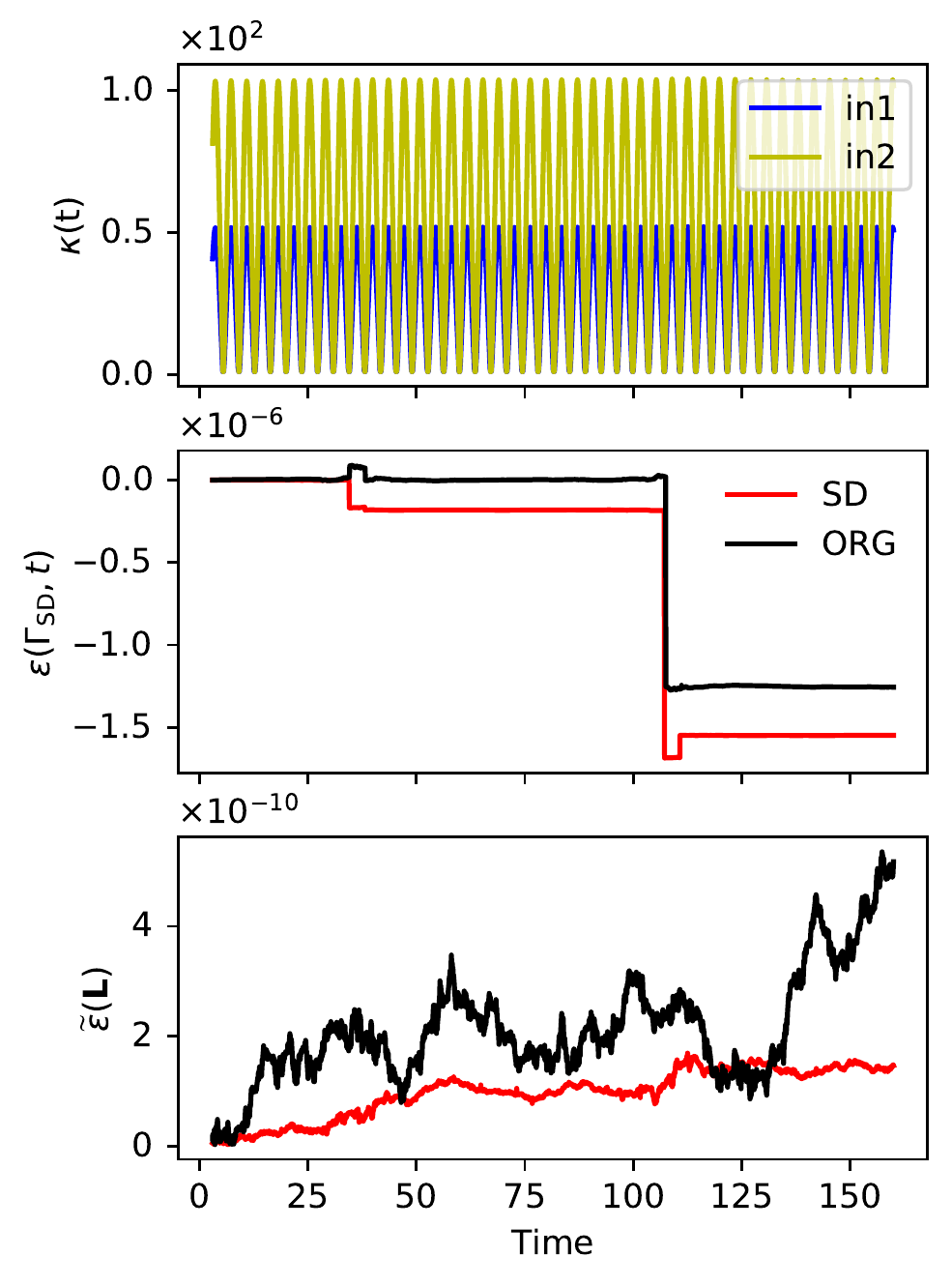}
  \caption{The evolution of $\kappa$, $\egsd$ and $\elv$ for the B-B system. The upper panel shows $\ki$ of the two inner binaries. The middle and lower panels compared $\egsd$ and $\elv$ for the SD and ORG cases.}
  \label{fig:bbhsd}
\end{figure}

On the other hand, since $a$ of two inner binaries do not evolve significantly, it is easy to observe the difference caused by the orbital resonance between the SD and the ORG method.
Although the SD method cannot keep the original period ratio of the two binaries, the mean motion resonance still exist because $\ki$ of the two binaries are related by Eq.~\ref{eq:kappa}.
The upper panel of Fig.~\ref{fig:bbhsd} shows the variation of $\ki$, where $\kib$ is twice of $\kia$.
Therefore, the period ratio changes from $1:1$ to $1:2$.
Since the orbits are slowed down, the resonance becomes stronger, which results in the larger drift in the SD case.

$\egsd$ keeps the order of $10^{-7}$.
Both SD and ORG methods show jumps when $\eia$ passes the maximum value due to the large numerical error at the pericenter of the first inner binary.
Such error can be reduced with a smaller $\Delta s$ (not shown here).
The behaviour of $|\Lv|$ is independent of $\eia$ and its relative error has an order of $10^{-10}$.

\subsection{Hyperbolic encounter (HB-B)}

\begin{table*}
  \caption{The initial condition of the system of a hyperbolic encounter between two binaries (HB-B). $\Porg$ of the outer orbit is the time to reach the pericenter. The inner binary orbits are the same as in the B-B system except in the ORG models, $\Einb$ uses three values for testing the impact of initial orbital phases. The SD model (SD-3) chooses the third value of $\Einb$.}
  \begin{tabular}{ccccc ccccc}
    \hline
          & $\mpr$ & $\ms$ & $a$ & $e$ & $\theta$ & $\phi$ & $\psi$ & $E$ & $\Porg$ \\
    \hline
     in1 & 0.900 & 0.100 & 0.00100 & 0.900 & 1.50 & 0.00 & 0.00 & 3.14 & $1.97\times10^{-4}$ \\
     in2 & 0.00900 & 0.00100 & 0.00200 & 0.900 & 1.50 & 0.00 & 0.00 & 3.00, 3.50, 3.14 & $5.62\times10^{-3}$ \\
     out & 1.00 & 0.0100 & -1.00 & 1.02500 & 0.100 & 0.00 & 0.00 & -2.00 & 1.71 \\
    \hline
  \end{tabular}
  \label{tab:hb-b}
\end{table*}

In star clusters, hyperbolic encounters between stars and binaries are frequent.
Thus we investigate whether the slow-down method can provide an acceptable result in such case.
The encounter happens in a short time interval.
When a perturber is far away, the binary has a weak perturbation so that $\kappa$ is large.
Once the encounter starts, $\kappa$ drops fast.
Thus, it is necessary to limit the change rate of $\kappa$, for example, by using the timescale criterion (Eq.~\ref{eq:kmaxt}).

We test a hyperbolic encounter between a massive binary and a low-mass binary (HB-B).
The initial condition is shown in Table~\ref{tab:hb-b}.
The mass ratio between the two inner binaries is $100$.
The hyperbolic (outer) orbit has an initial separation of $2.86$ and a pericenter distance of $0.0250$ ($25$ times of $\aia$).
Initially $\Eout=-2.00$ so that the time to reach the pericenter is about $1.71$.
Based on Eq.~\ref{eq:kappa}, $\kia$ has the maximum value of about $3\times10^4$ and can reaches the minimum limit of $1.0$.
Thus, $\kia$ varies about four order of magnitude in one encounter.
By using the timescale criterion with $c=0.1$ in Eq.~\ref{eq:kmaxt}, $\kia$ is limited to about $10^3$, which is one magnitude smaller.

Due to the high mass ratio, the second inner binary feel a strong perturbation from the first inner binary.
The final fate of the second binary depends on its orbital phase during the encounter.
In the ORG case, three values of $\Einb$ are used to test the effect of orbital phases.
The SD model use the third value of $\Einb$ listed in Table~\ref{tab:hb-b}.

\begin{figure*}
  \centering
  \includegraphics[width=0.8\textwidth]{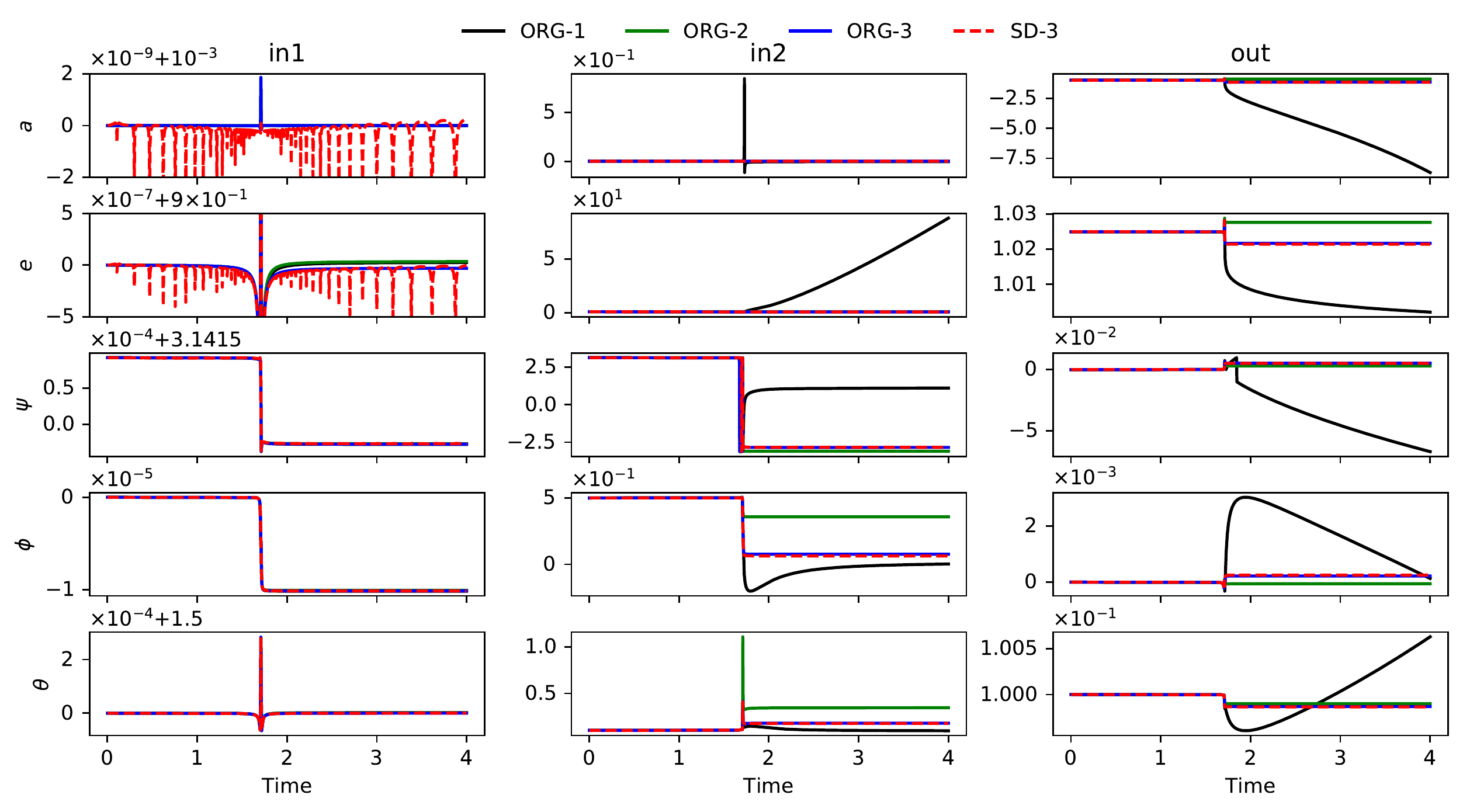}
  \caption{The evolution of orbital parameters of the two inner binaries and the outer encounter for the HB-B system.
    The solid and dashed lines represent the ORG and SD methods separately.
    Colors indicate different $\Einb$ in the ORG cases.
    The SD-3 and ORG-3 have the same initial $\Einb$.
    The plotting style is similar to Fig.~\ref{fig:bs}.}
  \label{fig:hbb}
\end{figure*}

The result of orbital revolution is shown in Fig.~\ref{fig:hbb}.
After the encounter, a jump appears in all orbital parameters.
The initial $\Einb$ significantly influences the final orbit of the secondary inner binary (a large divergence appears).
However, the SD and ORG method (SD-3 and ORG-3) can provide a similar result when the initial $\Einb$ are same ($3.14$).
Because the SD method loses the real orbital phase of the first binary, the result suggests that although the first binary is much more massive, its orbital phase has a minor impact on the evolution of the second binary.
Therefore, the SD method can provide an acceptable result.

\begin{figure}
  \centering
  \includegraphics[width=0.8\columnwidth]{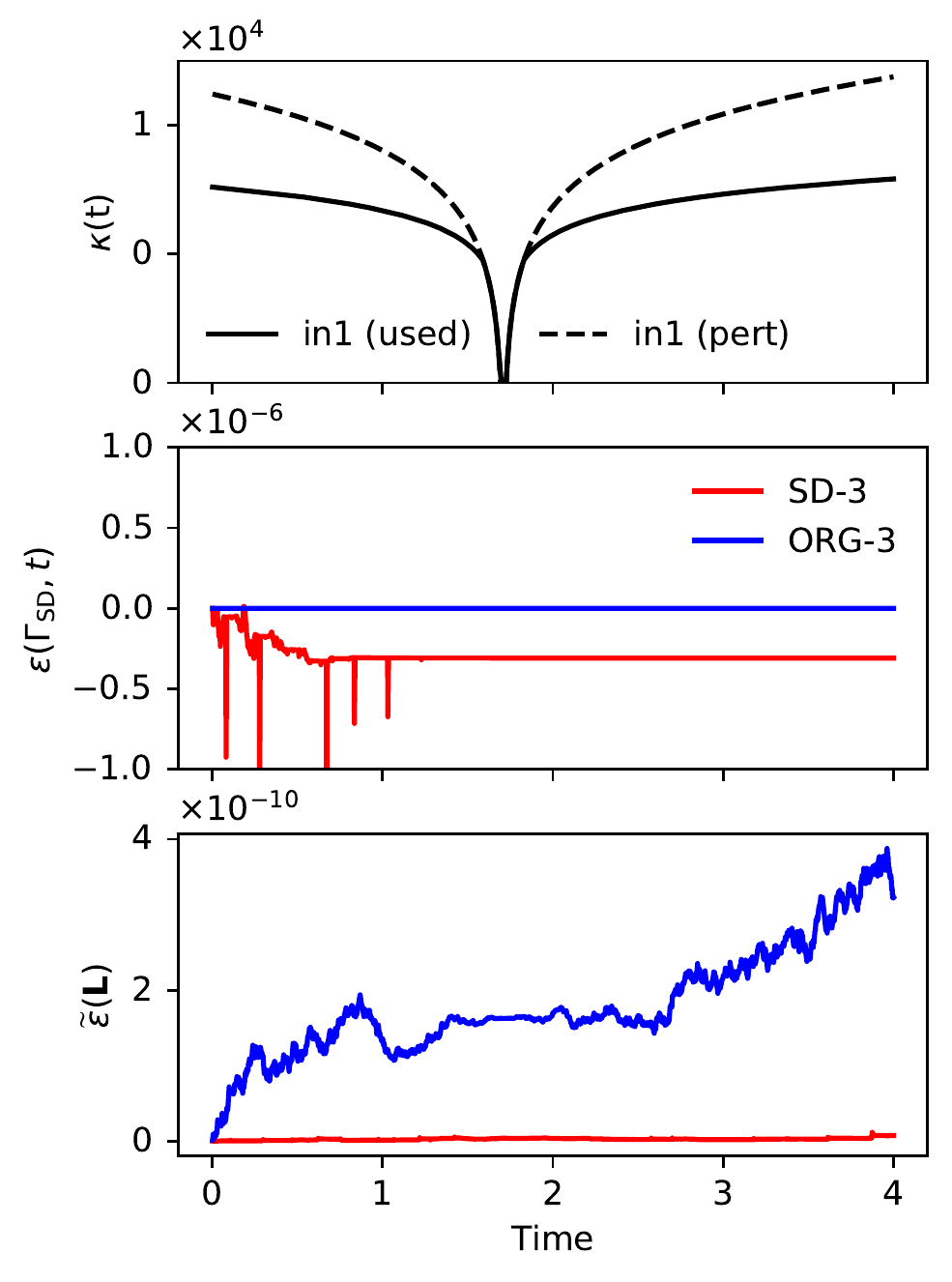}
  \caption{The evolution of $\kia$ (SD-3), $\egsd$ and $\elv$ for the HB-B system.
    In the upper panel, the dashed curve (used) is calculated by both the perturbation and the timescale criterion and the solid curve (pert) only applies the perturbation criterion.
    The SD-3 and ORG-3 models are compared for $\egsd$ and $\elv$ with a similar plotting style to that of Fig.~\ref{fig:bbhsd}.}
  \label{fig:hbbhsd}
\end{figure}

The upper panel of \ref{fig:hbbhsd} compared $\kia$ (for the SD-3 model) calculated by the perturbation criterion (Eq.~\ref{eq:kappa}) and by both the perturbation and the timescale criterion.
The timescale criterion ensures that $\kia$ changes more smoothly during the encounter.
$\ehsd$ has a larger error of  ($10^{-5}$) compared to $\eh$ at the beginning because $\kia$ is large.
It would be worse if the timescale criterion is not applied. 
Although $\egsd$ is not small, $|\Lv|$ is still well conserved (error has an order of $10^{-10}$), which again suggests that the conservation of $\Lv$ is independent of $\kappa$.

With $\Delta s \approx 1.58\times 10^{-6}$, the ORG method requires about $2.3\times10^{8}$ steps to reach $t=4$ while the SD method only needs $3.8\times 10^{6}$ steps.
Thus, the SD method provides a $60$ times faster performance.
Notice here we only evolve the system in a short time interval around the time of encounter.
In a star cluster, most of the time the binaries are weakly perturbed, thus averagely $\ki \gg 10^3$.
Therefore, the total integration steps of binaries are significantly reduced during the long-term evolution.
This example suggests that in the simulation of star clusters with many binaries, the SD method can dramatically improves the performance while the statistical result of encounters can be well reproduced.

\section{Discussion and conclusion}
\label{sec:conclusion}

In this work, we mathematically and numerically describe the slow-down time-transformed explicit symplectic (SDAR) method.
It combines the benefit of the symplectic integrator which conserves the Hamiltonian and angular momentum and the high efficiency of the slow-down method to handle the long-term evolution of hierarchical systems and close encounters.
An implementation of the method written in the c++ language, \code, is publicly available.
The code modularized for easily plugging in other $N$-body codes.

The Hamiltonian and the corresponding equation of motion are discussed in detail.
Although the physical energy ($H$) is not conserved, the method can provide a correct secular evolution.
We mathematically prove that $\Lv$ is always conserved with the slow-down method.
We also discussed how to measure the numerical error, $\egsd$, in the absence of energy conservation.

We show that in the LogH method, for a Kepler orbit the integration step $\diff s$ has the geometric meaning of eccentric anomaly(Eq.~\ref{eq:dsLdE}).
Using this feature, we can determine the number of steps per orbit by setting $ds$ using Eq.~\ref{eq:dsLdE}.
This is very useful for controlling the integration error.

On the other hand, when the LogH method is implemented in a hybrid $N$-body integrator, it is necessary to ensure that the integration can stop at a given time, in order to calculate the interactions between the members of few-body systems and the global system.
However, because time is an integrated value, before the integration finishes, the next time is unknown.
A few iteration of integration is needed to converge the next time to a certain value with a given limit of error.
Such iteration can be expensive if the time synchronization is frequent.
Thus, in order to avoid too many synchronization requests, it is necessary to design a good criterion for determination of the subsystems that need to be handled by the SDAR method.
Especially, the number of iteration steps should be much less than the number of integration steps.

By using the \ttl~code, we show three numerical examples (B-S, B-B and HB-B) to indicate that the SDAR method can well reproduce the Kozai-Lidov oscillation and can give an acceptable result of a hyperbolic encounter between two binaries.
When the averaged value of $\kappa$ is large, the slow-down method can provide significant performance improvement.
Especially, for binaries with weak perturbations, the method can provide a few order of magnitude less integration steps.
Thus, by combining this algorithm in an hybrid integrator for simulating a large $N$-body system, it is expected that a large fraction of binaries and hierarchical systems can be efficiently handled.
Such a code will be available soon in our following-up work.

\section*{Acknowledgments}

L.W. thanks the financial support from JSPS International Research Fellow (School of Science, The university of Tokyo).

\appendix

\section{Proof for the conservation of angular momentum }
Here we provide the proof for the conservation of angular momentum in the slow-down method.
First, the conservation of angular momentum for an isolated binary can be described as
\begin{equation}
  \begin{aligned}
    &\{\Lb, \Hb\}  = \bm{0}  \\
    &\Lb = \ma \ra \times \va + \mb \rb \times \vb \\
    &\Hb = \frac{1}{2}\ma \va^2 + \frac{1}{2} \mb \vb^2 - \frac{G \ma \mb}{|\ra-\rb|}\\
  \end{aligned}
  \label{eq:LH}
\end{equation}
where $\Lb$, $\Hb$ are the angular momentum and Hamiltonian of a binary system in the center-of-the-mass frame.

After a slow-down factor $\kappa(t)$ is included,
\begin{equation}
  \{\Lb,\Hsd\} = \{\Lb, \frac{1}{\kappa} \Hb \} = \frac{1}{\kappa}   \{\Lb, \Hb\} = \bm{0}.
\end{equation}
$\Lb$ is still conserved.
When the center-of-the-mass term is included,
\begin{equation}
  \begin{aligned}
    \Hsd = & \frac{1}{\kappa} \left [\frac{1}{2} \ma (\va-\vcm)^2 + \frac{1}{2} \mb (\vb-\vcm)^2  - \frac{G \ma \mb}{|\ra-\rb|} \right ] \\
    & + \frac{1}{2} (\ma+\mb) \vcm^2,
  \end{aligned}
\end{equation}
it can be shown that the conservation (Eq.~\ref{eq:LH}) still exists.

If a new body is added to form a triple, the additional term appears in the angular momentum:
\begin{equation}
  \Lv = \Lb + \Lc.
\end{equation}
Then
\begin{equation}
  \{\Lv, \Hb\} = \{\Lb, \Hb\} + \{\Lc, \Hb\} = \{\Lc, \Hb\}.
\end{equation}
Since $\Lc$ is independent of binary position and velocity,
\begin{equation}
  \{\Lv, \Hb\} = \{\Lc, \Hb\} = \bm{0}.
\end{equation}

The additional term to $\Hsd$ from the third body is
\begin{equation}
  \Hc = \frac{1}{2} \mc \vc^2 - \sum_i^{2} {\frac{G \mi \mc}{|\ric|}}.
  \label{eq:Hc}
\end{equation}
It can be proved that 
\begin{equation}
  \{\Lv, \Hc\} = \bm{0}.
\end{equation}
Thus,
\begin{equation}
  \{\Lv, \Hsd\} = \{\Lv, \Hb\} + \{\Lv, \Hc\} = \bm{0}.
  \label{eq:LH3}
\end{equation}
The angular momentum is conserved for the triple.

For systems with one binary and many singles, $4^{th}, 5^{th}...$ particles can be added one by one, and for the $k^{th}$ particle, the additional Hamiltonian term is
\begin{equation}
  \Hi = \frac{1}{2} \mi \vi^2 - \sum_j^{i-1} {\frac{G \mi \mj}{|\rij|}}.
\end{equation}
Since the second term in $\Hi$ is a linear combination of $k-1$ pair potential energy, it is not difficult to show that
\begin{equation}
  \{\Lv, \Hi\} = \bm{0}.
\end{equation}
Eventually,
\begin{equation}
  \{\Lv, \Hsd\} = \bm{0}
\end{equation}
Thus, $\Lv$ is conserved in such system.

On the other hand, here we add many singles to a slow-down binary and indicate that $\Lv$ is conserved.
We can consider this process in an opposite way: add one slow-down binary to a systems of many singles, the new $\Lv$ is still conserved.
Thus, it can be proved that adding arbitrary slow-down binaries to the systems, $\Lv$ is always conserved.

\label{lastpage}

\end{document}